\documentclass[times,twocolumn,final,nopreprintline]{elsarticle}

\usepackage{medima2}
\usepackage{framed,multirow}
\usepackage{amsmath}

\usepackage{amssymb}
\usepackage{latexsym}
\usepackage{array}
\usepackage{graphicx}
\usepackage{subcaption}
\usepackage{comment}

\usepackage{url}
\usepackage{xcolor}
\usepackage{hyperref}

\definecolor{newcolor}{rgb}{.8,.349,.1}

\begin{document}

\verso{Chayan Banerjee \textit{et~al.}}

\begin{frontmatter}

\title{PINNs for Medical Image Analysis: A Survey}%

\author[1]{Chayan \snm{Banerjee}}
\ead{c.banerjee@qut.edu.au}
\author[1]{Kien \snm{Nguyen}\corref{cor1}}
\cortext[cor1]{Corresponding author}
\ead{k.nguyenthanh@qut.edu.au}
\author[2]{Olivier \snm{Salvado}}
\ead{olivier.salvado@csiro.au}
\author[3]{Truyen \snm{Tran}}
\ead{truyen.tran@deakin.edu.au}
\author[1]{Clinton \snm{Fookes}}
\ead{c.fookesb@qut.edu.au}

\address[1]{Queensland University of Technology (QUT), 2 George St, Brisbane QLD 4000, Australia}
\address[2]{Commonwealth Scientific and Industrial Research Organisation (CSIRO), 1 Technology Court, Pullenvale QLD 4069, Australia}
\address[3]{Deakin University, 75 Pigdons Rd, Waurn Ponds, VIC 3216, Australia}


\begin{abstract}
The incorporation of physical information in machine learning frameworks is transforming medical image analysis (MIA). By integrating fundamental knowledge and governing physical laws, these models achieve enhanced robustness and interpretability. In this work, we explore the utility of physics-informed approaches for MIA (PIMIA) tasks such as registration, generation, classification, and reconstruction. We present a systematic literature review of over 80 papers on physics-informed methods dedicated to MIA. We propose a unified taxonomy to investigate what physics knowledge and processes are modelled, how they are represented, and the strategies to incorporate them into MIA models. We delve deep into a wide range of image analysis tasks, from imaging, generation, prediction, inverse imaging (super-resolution and reconstruction), registration, and image analysis (segmentation and classification). For each task, we thoroughly examine and present in a tabular format the central physics-guided operation, the region of interest (with respect to human anatomy), the corresponding imaging modality, the dataset used for model training, the deep network architecture employed, and the primary physical process, equation, or principle utilized. Additionally, we also introduce a novel metric to compare the performance of PIMIA methods across different tasks and datasets. Based on this review, we summarize and distil our perspectives on the challenges, open research questions, and directions for future research. We highlight key open challenges in PIMIA, including selecting suitable physics priors and establishing a standardized benchmarking platform.
\end{abstract}


\end{frontmatter}



\section{Introduction}

Recent advancements in MIA have achieved exceptional performance in tasks like image registration, image generation, feature extraction, image classification and image reconstruction \cite{DLMedical}. These tasks are essential for aiding in diagnosis, treatment planning, and facilitating therapeutic interventions. Yet, these achievements often rely on complex, data-intensive models lacking robustness, interpretability, and alignment with governing physical rules and commonsense reasoning \cite{XAIMedical,AdvAttacksMedical}.

Applying generic machine learning approaches to medical image analysis presents unique challenges arising from the scarcity of data, the expense of data collection, rigorous requirements for interpretability, precision standards, and considerable inter-patient variability. Purely data-driven models demand a large quantity of training data, exhibit slow convergence, and often require large model architectures with millions of parameters to train. In the MIA context due to the scarcity of quality data, all these issues become more prominent and limit the efficacy of data-driven machine learning models.

Physics equations and mathematical models can provide simplified analogies for understanding physical processes inside human anatomy and physiological mechanisms of imaging to facilitate medical image analysis tasks, offering valuable insights even though they do not precisely mirror the complexity of these systems.
For example, modelling blood flow as a non-newtonian fluid \cite{toghraie2020blood} or use of Navier-Stokes equations as the foundational physics model for understanding and simulating cardiovascular hemodynamics \cite{li2021prediction}. 
Additionally, medical image acquisition can be formulated as differential equations, boundary conditions and constraints. For example, dynamics of CT perfusion scanning \cite{de2023spatio},  MRI image acquisition physics \cite{borges2024acquisition} and MRI hardware constraints \cite{peng2022learning}.
%
Physics-based methods, grounded in fundamental equations and domain knowledge, promise improved reliability and system safety by embodying the actual physical relationships at play \cite{balageas2010structural,hu2019case}, suggesting a need for a paradigm shift towards incorporating physical laws in MIA.

\begin{figure}[h!]
    \centering
    \includegraphics[width=0.99\columnwidth]{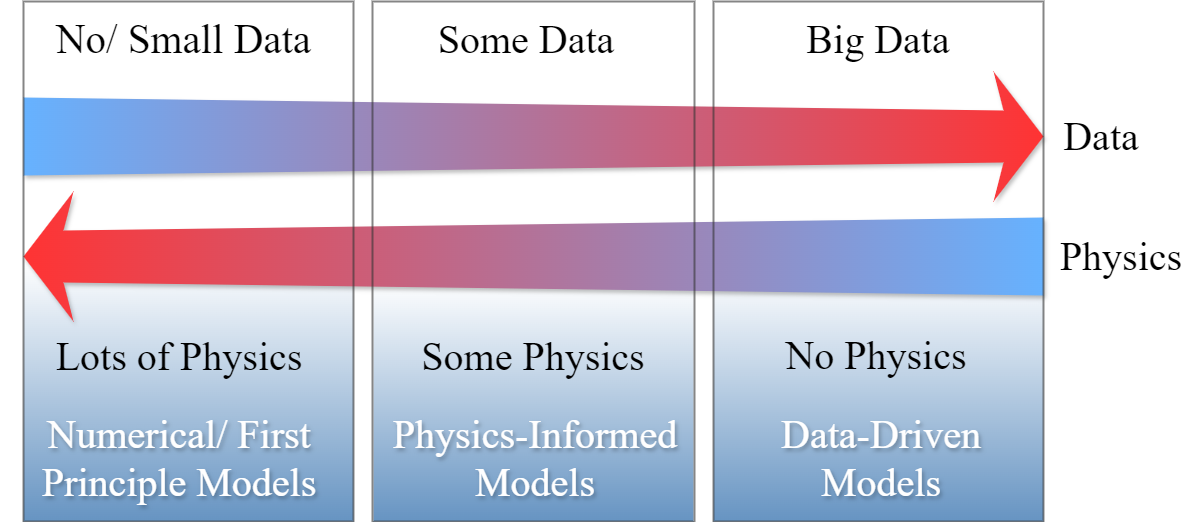}
    \caption{Physics-informed machine models leverage physics information to enhance efficiency and plausibility, bridging data-driven and numerical models, vital for accurate representations across medical domains.}
    \label{fig:PIMLbenefits}
\end{figure}

Recent studies highlight the advantages of incorporating physics principles with machine learning, establishing a dominant paradigm in the field. Physics-informed machine learning (PIML), which integrates mathematical physics into machine learning models, enhances solution relevance and efficiency. As illustrated in Fig.~\ref{fig:PIMLbenefits}, physics-informed models provide a balance between numerical models and purely data-driven models with regard to the amount of data and the amount of physics. PIML approaches accelerate neural network training, improve model generalization with fewer data, and manage complex applications while ensuring solutions adhere to physical laws \cite{karniadakis2021physics,hao2022physics}.
Incorporating physical principles into machine learning, as seen in PIML approaches, significantly boosts the robustness, accuracy, efficiency, and functionality of computer vision models \cite{hao2022physics,meng2022physics,karniadakis2021physics}. As MIA is progressively integrating machine learning and deep learning techniques to enhance its capabilities, it can exhibit the same benefits that PIML can bring. 


Besides the similarity with generic machine learning and computer vision tasks, MIA also presents distinct challenges due to the nature of its input data, which primarily consists of complex and noisy medical images of human anatomy. These challenges include limited and high-cost data, stringent interpretability requirements, high precision demands, and significant inter-patient variability. Medical images capture intricate phenomena occurring within the human body, such as blood flow, abnormalities like tumours, metal artifacts, and dynamic organ activity such as the beating heart and lung respiration. They are generated through specialized imaging technologies and often contain machine-specific information reflecting the specific processing steps taken by healthcare practitioners. Effectively interpreting medical images necessitates advanced domain-specific expertise in medical physics, biomedical engineering, radiology, and biomechanics. This multidisciplinary knowledge is essential to address the intricate complexities inherent in biological tissue characteristics and the sophisticated techniques used in medical imaging modalities, such as MRI, CT, and ultrasound. These unique characteristics distinguish MIA from the analysis of generic images.
This distinction underlines the need for models and methods specifically designed for MIA tasks, and leverages the unique physics information available, leading to the exploration of the PIMIA field.
The paper reviews state-of-the-art physics-informed strategies in MIA, focusing on how physics knowledge is integrated into algorithms, what the physical processes are modelled as priors, and the specialized network architectures or augmentations employed to weave in physics insights.

\begin{figure}[h!]
    \centering
    \begin{subfigure}[b]{1\linewidth}
        \includegraphics[width=0.99\linewidth]{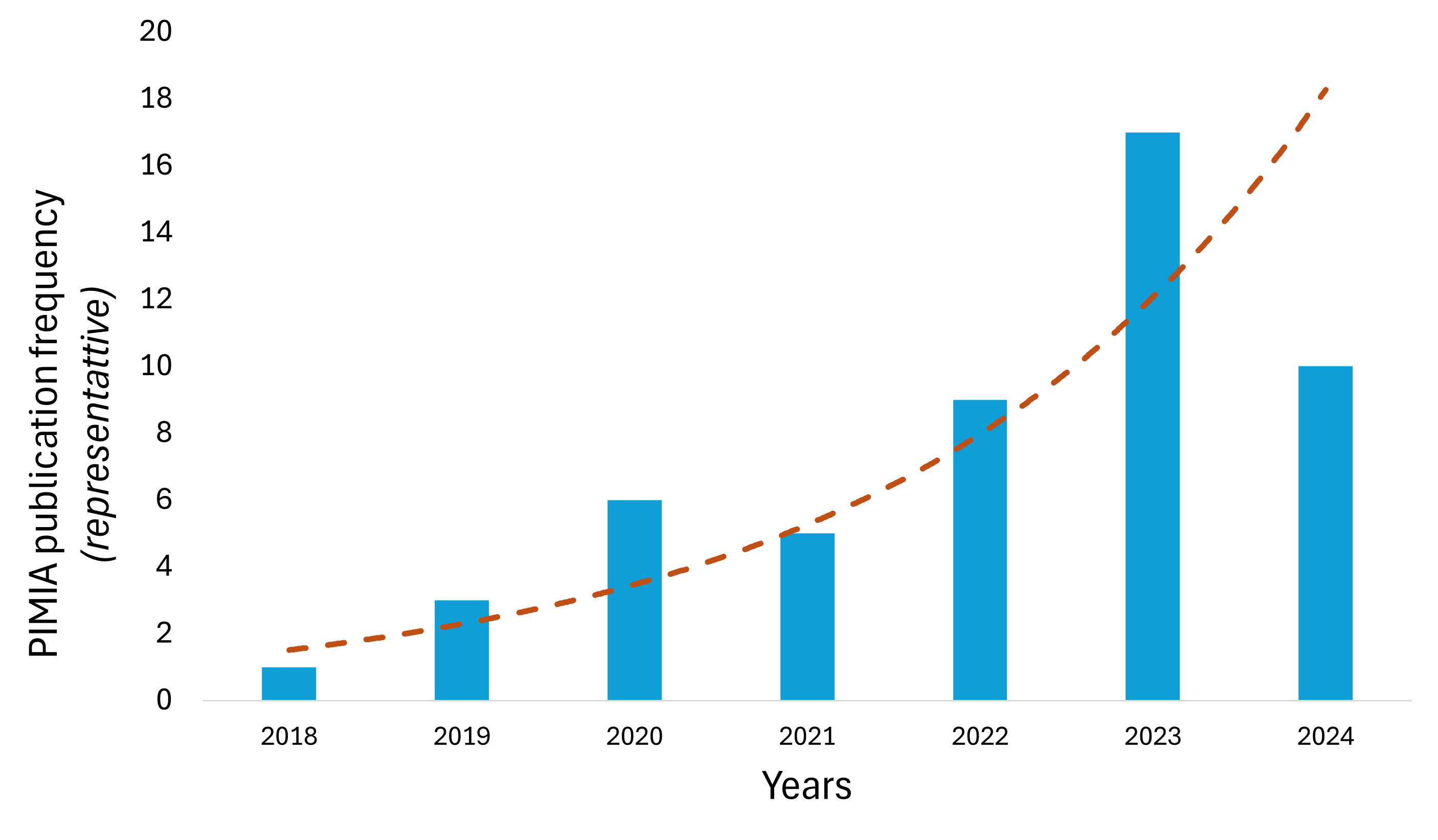}
        \caption{PIML research over the years ( 2018 - 2024 (\textit{till June}) ), clearly showing an exponential trend (shown in dashed line).}
        \label{fig:trend}
    \end{subfigure}
   \hfill
    \begin{subfigure}[b]{1\linewidth}
        \includegraphics[width=\linewidth]{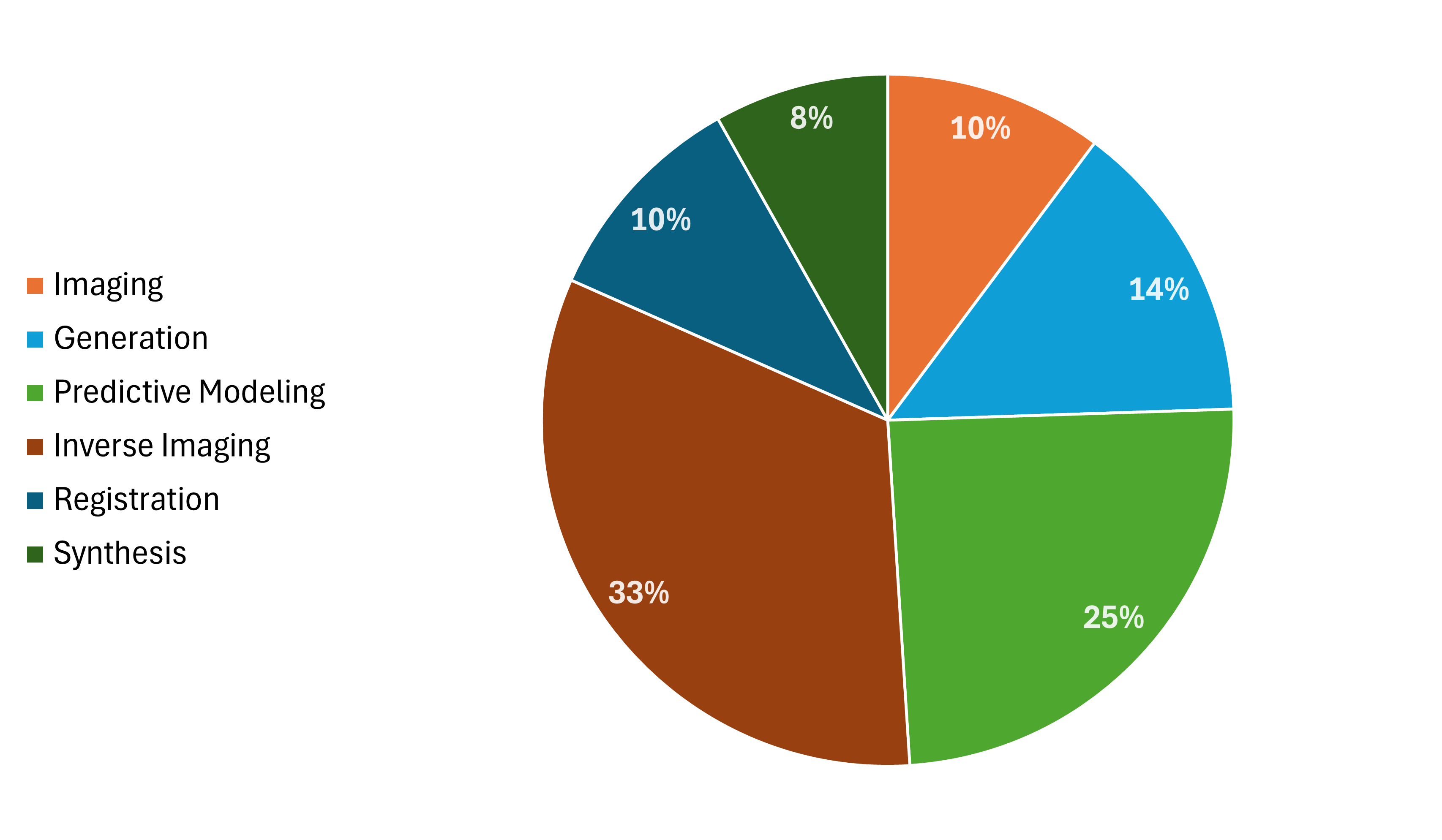}
       \caption{Research share of different PIMIA tasks, from the surveyed papers. }
       \label{fig:share}
    \end{subfigure}
    \caption{Plots showing the latest research trend on the topic of PIMIA, (a) Published work over the years and (b) Research share of PIMIA tasks.}
    \label{fig:trend-n-share}
\end{figure}

The recent literature on PIMIA has shown an exponentially increasing trend as evident through the plethora of publications over the last 7 years, see Fig.~\ref{fig:trend}.
The accompanying pie chart presents the research share w.r.t. different PIMIA tasks. Clearly showing a heightened focus on physics incorporation in PIMIA tasks like inverse imaging especially image reconstruction and predictive modeling.

Our contributions in this paper are summarized as follows:
\begin{itemize}
    \item We propose a unified taxonomy to investigate what physics knowledge/processes are modelled, how they are represented, and the strategies to incorporate them into MIA models.
    \item We delve into a wide range of image analysis tasks, from imaging, generation, prediction, inverse imaging (super-resolution and reconstruction), registration, and image analysis (segmentation and classification)
    \item For each task, we thoroughly examine and present in an easily referenceable tabular format the central physics-guided operation, the region of interest (with respect to human anatomy), the corresponding imaging modality, the dataset used for model training, the deep network architecture employed, and finally, the primary physical process, equation, or principle utilized.

    \item Based on the review of tasks, we summarize our perspectives on the challenges, open research questions, and directions for future research. 

We highlight key open challenges in PIMIA, including selecting suitable physics priors and establishing a standardized benchmarking platform. While physics may have been integrated into the training pipeline, particularly through data augmentation, tasks such as classification, segmentation, and super-resolution have yet to fully leverage physics priors. These tasks are ranked in order of least to moderately efficient utilization of physics information.
 
\end{itemize}

\noindent
\textbf{Focus of the current survey}:\newline
The field of physics-informed machine learning (PIML) is rapidly expanding, highlighted by surveys across various domains \cite{hao2022physics} including cyber-physical systems \cite{rai2020driven}, hydrology \cite{zhang2019recent}, fluid mechanics \cite{cai2022physics}, climate modelling \cite{kashinath2021physics} and reinforcement learning \cite{banerjee2023survey}.

Several surveys have examined the integration of physics and knowledge in MIA, often highlighting specific areas. These surveys provide valuable insights, yet there remains an opportunity for a more comprehensive overview that captures broader trends and advancements in the field. \cite{banerjee2023physics} conducted a comprehensive survey on the incorporation of physics in computer vision tasks, albeit with a broader focus on computer vision in general, lacking detailed coverage of MIA. Conversely, \cite{liu2021anatomy} concentrated on anatomy-aided deep learning (DL) specifically for medical image segmentation, presenting a thorough examination of anatomical information categories and representation methods. \cite{hammernik2023physics} provided an in-depth survey of physics-driven DL techniques for MRI reconstruction, encompassing both linear and nonlinear forward models as well as classical and DL approaches. \cite{xia2023physics} scrutinized physics- and model-based data-driven methods tailored for low-dose computed tomography (LDCT), emphasizing the integration of imaging physics principles into deep network architectures. In another work \cite{maier2022known}, the author discusses ``known operator learning'' in medical imaging.

\textcolor{black}{Unlike existing surveys, our study offers a comprehensive review of approaches that integrate physics-based information into MIA (i.e PIMIA). While other reviews either concentrate on non-physics-specific methods or focus on specific imaging modalities, our study explores a diverse range of physics-based approaches and their applications across various imaging modalities and regions of interest. Additionally, research in the realm of PIML adapted for MIA is relatively recent, significantly rising over the past 3-4 years, despite the long-standing presence of knowledge-guided approaches in MIA.}

%

\noindent\textbf{Inclusion and Exclusion Criteria for Article Selection:}
\textcolor{black}{Our selection criteria are guided by several key considerations to ensure the relevance and quality of the literature we review. First, we prioritize works that are closely aligned with the core topics of our survey, particularly those that incorporate physics-based approaches rather than methods guided by \textit{domain-specific knowledge} and \textit{non-physics-specific information}. 
%
%
Additionally, while we generally focus on publications from peer-reviewed journals and conferences, we have made exceptions to include some pre-prints, recognizing the importance of capturing novel advancements in the field. Lastly, our review is strictly confined to literature pertaining to medical image analysis, ensuring that all included works are directly relevant to this specialized area.\newline
To ensure transparency, we have provide a detailed description of these criteria below. This includes the specific databases and journals searched, the keywords used, and the temporal scope of the included literature. By applying these criteria, we aim to provide a comprehensive and unbiased synthesis of the existing research, addressing the key themes and advancements in the field. }

\textcolor{black}{Data for this review was systematically collected from leading journals such as \textit{Medical Image Analysis}, \textit{Physics in Medicine and Biology}, \textit{IEEE Transactions on Medical Imaging}, \textit{Medical Physics}, \textit{Nature Reviews Physics}, and \textit{Nature Machine Intelligence}. Our search encompassed major databases including PubMed, IEEE Xplore, ScienceDirect, SpringerLink, Wiley Online Library, ACM Digital Library, ArXiv, Scopus, JSTOR, and Google Scholar.The keywords searched included \textit{physics informed medical image analysis} \textit{physics informed imaging}, \textit{physics informed generation}, \textit{physics incorporated predictive modeling}, \textit{physics informed inverse imaging}, and \textit{physics informed image registration}. The papers surveyed were specifically selected from the past 7 years due to the rapid advancements in physics informed approaches in medical imaging and analysis during this period. The time frame ensures that the review reflects the most current trends and breakthroughs in PIMIA, providing a relevant and up-to-date overview of the state-of-the-art research and practices.}

The paper is structured as follows: Section~\ref{sec: PIML} introduces Physics Informed Machine Learning (PIML), especially its benefits and categories to provide context for its application in MIA. Section~\ref{sec: PIMIA} talks about the PIMIA approach and introduces a taxonomy for physics priors. Section~\ref{sec:PIMIA_survey} presents the state-of-the-art of the conventional works published in the realm of PIMIA. In Section~\ref{sec:quant_study} we provide a quantitative study and provide insights regarding actual benefits and best-fit scenarios of using PI in MIA. In Section~\ref{sec:Challenges}
we discuss the existing challenges in incorporating physics priors into PIMIA tasks and highlight the research gaps and future research avenues. Finally in section \ref{conclusion} we conclude this article, where we highlight its crucial takeaways.

\begin{table}[h!]
\centering
\scalebox{0.7}{
\begin{tabular}{| p{0.3\columnwidth} | p{1.0\columnwidth} |}
\hline
\textbf{Abbreviation} & \textbf{Definition} \\
\hline
DECT & Dual-Energy Computed Tomography \\
\hline
GRE MRI & Gradient Echo Magnetic Resonance Imaging \\
\hline
CT & Computed Tomography \\
\hline
MRI & Magnetic Resonance Imaging \\
\hline
PET & Positron Emission Tomography \\
\hline
US & Ultrasound \\
\hline
DWI-MRI & Diffusion-Weighted Imaging Magnetic Resonance Imaging \\
\hline
4D-Flow MRI & Four-Dimensional Flow Magnetic Resonance Imaging \\
\hline
CBCT & Cone Beam Computed Tomography \\
\hline
MDCT & Multi-Detector Computed Tomography \\
\hline
MRV & Magnetic Resonance Venography \\
\hline
ECG & Electrocardiography \\
\hline
QSM & Quantitative Susceptibility Mapping \\
\hline
DSA & Digital Subtraction Angiography \\
\hline
ODT & Optical Diffraction Tomography \\
\hline
TRUS & Transrectal Ultrasound \\
\hline
CXR & Chest X-Ray \\
\hline
EEG & Electroencephalography \\
\hline
TCD US & Transcranial Doppler ultrasound.\\
\hline
SWI & Susceptibility weighted imaging \\
\hline
\end{tabular}
}
\caption{List of Abbreviations used in this article}
\label{table:abbreviations}
\end{table}

\section{Physics-informed Machine Learning (PIML)} \label{sec: PIML}
This section briefly introduces the physics-informed machine learning or PIML paradigm. It also discusses a taxonomy of biases typically used to categorize the point of incorporation of physics principles, data or equation w.r.t to a machine learning pipeline.

PIML endeavours to fuse mathematical physics models and observational data into the learning process to steer it towards physically consistent solutions in scenarios characterized by partial observation, uncertainty, and high dimensionality \cite{kashinath2021physics,hao2022physics,cuomo2022scientific}. The inclusion of physics information, which embodies the fundamental principles of the modelled process, enriches ML models by conferring significant advantages \cite{kashinath2021physics,meng2022physics}:

\begin{enumerate}
\item Helps ensure that the ML model is both physically and scientifically coherent.
\item Enhances data efficiency in model training, enabling training with fewer data.
\item Expedites the model training process, leading to faster convergence to an optimal solution.
\item Boosts the generalizability of trained models, enabling better predictions in unseen scenarios.
\item Enhances the transparency and interpretability of models, rendering them explainable and more trustworthy.
\end{enumerate}





\textcolor{black}{An important example of this approach of embedding physics information in the ML paradigm is the family of Physics-Informed Neural Networks (PINNs). PINNs are a class of deep learning algorithms that seamlessly integrate data and abstract mathematical operators, including partial differential equations (PDEs). This integration allows for more interpretable ML methods that remain robust despite imperfect data, such as missing or noisy values, and can provide accurate and physically consistent predictions. By leveraging both data and physical models, PINNs are particularly suited for solving forward and inverse problems, discovering hidden physics, and tackling high-dimensional problems in various scientific and engineering applications.}

\begin{figure}
    \centering
    \includegraphics[width=0.98\linewidth]{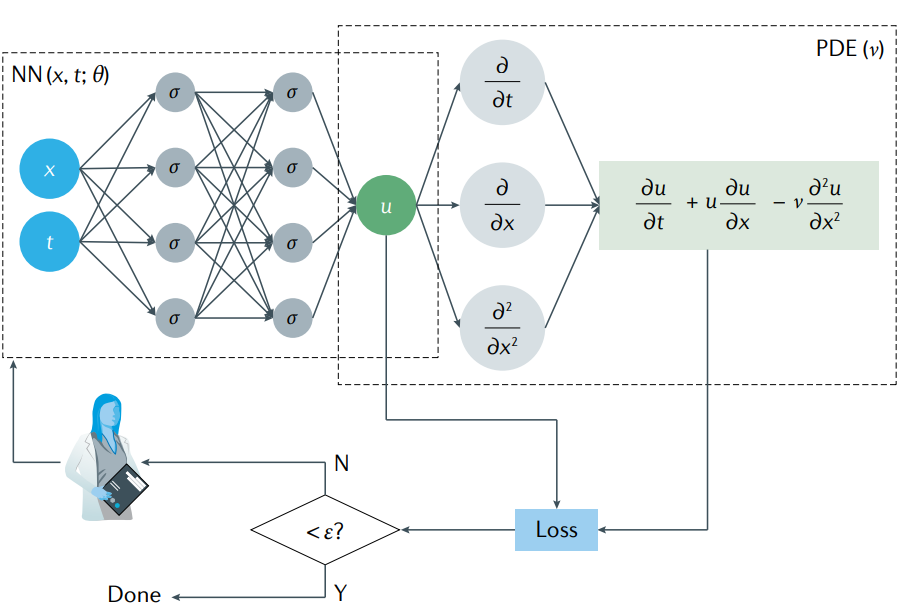}
    \caption{The PINN algorithm from \cite{karniadakis2021physics}}
    \label{fig:PINN_fig}
\end{figure}

\textcolor{black}{ We show an example of a typical PINN in Fig:~\ref{fig:PINN_fig} designed to approximate solutions to PDEs by leveraging neural networks. The PINN is defined by the neural network function NN\( (X, t; \theta) \), where \( X = (x, y, z) \) and \( t \) are the spatial and temporal coordinates, respectively, and \( \theta \) represents the trainable parameters. The network utilizes a nonlinear activation function 
$\sigma$. \newline
Measurement data is specified as \( \{(x_i, t_i, u_i)\} \), where \( x_i \) and \( t_i \) are spatial and temporal coordinates, respectively, and \( u_i \) are the observed values of \( U \) at these coordinates. Additionally, residual points \( \{(x_j, t_j)\} \) are defined, where the residuals of the PDE will be evaluated.\newline
Mathematically, the total loss function \( L \) is formulated as:
\[
L = \lambda_{\text{data}} L_{\text{data}} + \lambda_{\text{PDE}} L_{\text{PDE}}
\]
Here, \( \lambda_{\text{data}} \) and \( \lambda_{\text{PDE}} \) are weighting factors that balance the contributions of the data loss and the PDE loss.\newline
Training the neural network involves finding the optimal parameters \( \theta^* \) that minimize the total loss function \( L \):
\[
\theta^* = \arg \min_{\theta} L
\]
This optimization ensures that the neural network solution NN\((X, t; \theta) \) fits the observed data and adheres to the underlying physical laws described by the PDE. The algorithm runs iteratively until the loss is below a certain threshold $\epsilon$ (where $\epsilon$ is a positive number) or is terminated by the human operator.
}

\begin{figure}[t!]
    \centering
    \includegraphics[width=0.98\linewidth]{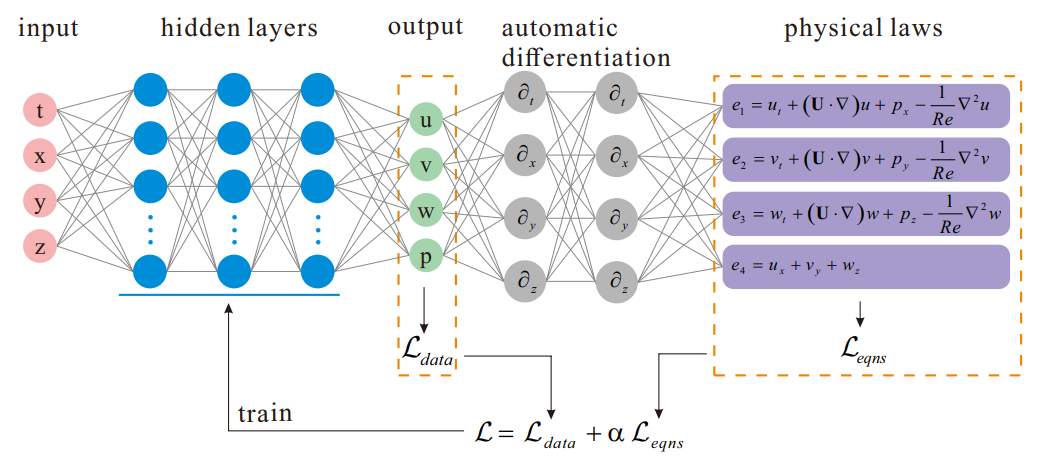}
    \caption{Example of PINN application from \cite{wang2022dense}}
    \label{fig:PINN_example}
\end{figure}

\textcolor{black}{Next we discuss the work by \cite{wang2022dense} from fluid-dynamics and  show how PINNs are used to enhance velocity field measurements from sparse data. In this example by minimizing a loss function based on sparse data and Navier–Stokes PDE equations, the PINN (see Fig~\ref{fig:PINN_example}) reconstructs dense velocity fields and predicts pressure, showcasing significant improvements in resolution and accuracy for flow visualization and analysis. The PINN is defined by the neural network \( NN(X, t; \theta) \), the network approximates both the velocity field \( U = (u, v, w) \) and the pressure \( p \). \newline
The loss function \( L \) for training the PINN is given by:
\[
L = L_{\text{data}} + \alpha L_{\text{eqns}},
\]
where \( L_{\text{data}} \) quantifies the deviation from experimental data and \( L_{\text{eqns}} \) enforces the physical laws described by the Navier-Stokes equations. The parameter \( \alpha \) adjusts the relative importance of the data fit and the PDE constraints. This approach aims to improve the resolution and accuracy of the reconstructed velocity fields and pressure distributions.}

\vspace{0.1cm}
\noindent Previous literature e.g. \cite{karniadakis2021physics} has identified three strategies for incorporating physics knowledge/priors into machine learning models: observational bias, learning bias, and inductive bias, these are discussed below, 

\subsection{Biases as taxonomy in PIML}
\textit{Observational bias:} 
\textcolor{black}{Observational data are foundational inputs in machine learning, essential for capturing and reflecting underlying physical principles across diverse scales and modalities \cite{karniadakis2021physics}. This data is instrumental in imparting knowledge of fundamental physical laws and principles to ML models \cite{lu2021learning,kashefi2021point,li2020fourier,yang2019conditional}. Sources of observational data include sensor networks (environmental, seismic), remote sensing (satellite imagery, weather data), experimental measurements (laboratory, wind tunnels), and high-fidelity simulations (e.g., Computational Fluid Dynamics (CFD), Finite Element Analysis (FEA)). These diverse sources capture intricate phenomena across different spatial and temporal scales, providing rich insights into physical systems. Incorporating such data during the training phase enables Deep Neural Networks (DNNs) to effectively model underlying physical processes. Variable fidelity observations ensure comprehensive coverage of physical behaviors, thereby enhancing the model's capacity for accurate interpolation and adherence to physical symmetries and conservation laws.}

\textit{Learning bias:} This approach enforces prior knowledge/physics information through soft penalty constraints. Methods in this category augment loss functions with additional terms based on the physics of the underlying process, such as momentum and conservation of mass. For instance, physics-informed neural networks (PINN) integrate information from both measurements and partial differential equations (PDEs) by embedding the PDEs into the loss function of a neural network using automatic differentiation \cite{karniadakis2021physics}. 
Some notable examples of soft penalty-based approaches include statistically constrained GAN \cite{wu2020enforcing}, PI auto-encoders \cite{erichson2019physics}, and encoding invariances by soft constraints in the loss function InvNet \cite{shah2019encoding}. 

\textit{Inductive biases:} Prior knowledge can be incorporated through custom neural network-induced 'hard' constraints.
For example, Hamiltonian NN \cite{greydanus2019hamiltonian} encodes better inductive biases into NNs, drawing inspiration from Hamiltonian mechanics, and training models to respect exact conservation laws. Cranmer et al. introduced Lagrangian Neural Networks (LNNs) \cite{cranmer2020lagrangian}, which can parameterize arbitrary Lagrangians using neural networks, and unlike most HNNs, LNNs can work where canonical momenta are unknown or difficult to compute.
Additionally, tensor basis networks (TNNs) \cite{ling2016reynolds} incorporate tensor algebra into their operations and structure, allowing them to exploit the high-dimensional structure of tensor data more effectively than traditional neural networks.
\cite{meng2022learning} employs a Bayesian framework where functional priors are learned using a PI-GAN from data and physics. This is followed by using the Hamiltonian Monte Carlo (HMC) method to estimate the posterior PI-GAN's latent space. 

MIA is progressively integrating machine learning (ML) and deep learning (DL) techniques to enhance its capabilities. PIML has demonstrated significant promise in augmenting the efficacy of MIA \cite{banerjee2023physics,hammernik2023physics,xia2023physics}. Subsequent sections will elucidate the application methodologies of PIML within the domain of MIA.

\section{Physics-Informed Medical Image Analysis (PIMIA)} \label{sec: PIMIA}
Physics-informed medical image analysis refers to a computational approach that incorporates principles from physics into the analysis of images, specific to medical discipline. PIMIA incorporates concepts and methodologies that are used in physics-informed machine learning (PIML), which is a broader term encompassing various domains, data types and modalities of physics information. In medical imaging, various physical processes govern how images are formed and how tissues interact with imaging modalities such as X-rays, MRI, CT scans, and ultrasound. The physics-informed analysis leverages this understanding to improve image reconstruction, segmentation, registration, and other tasks.

This approach typically involves the development of mathematical models that describe the physical phenomena underlying image formation, acquisition, device characteristics, and tissue properties. These models are then integrated into image analysis algorithms to enhance the accuracy, robustness, and interpretability of medical image processing tasks. Additionally, physics-informed methods can provide more reliable and clinically relevant results compared to purely data-driven approaches by constraining the solutions.

\subsection{Physics incorporation in general ML and MIA:}
In both general machine learning approaches and MIA, the incorporation of physics can enhance model performance and interpretability. However, there are notable differences in how physics is integrated into these two domains. In general machine learning, physics incorporation often involve the development of physics-inspired models or the inclusion of physical constraints as regularization terms in optimization objectives. For example, in fluid dynamics simulations, machine learning models may incorporate principles of conservation laws and fluid dynamics equations to improve predictions. 

On the other hand, in MIA, physics incorporation is more directly tied to the underlying imaging processes and tissue properties. For instance, in MRI reconstruction, physics-based models describing the signal formation process, such as the Fourier transform and T1/T2 relaxation processes, are integrated into reconstruction algorithms to enhance image quality and reduce artifacts. Moreover, PIMIA often involves the calibration of imaging systems to account for physical effects like attenuation and scattering in X-ray or ultrasound imaging, ensuring more accurate quantitative measurements and diagnostic assessments. Therefore, while both domains benefit from physics incorporation, the emphasis and application differ, reflecting the specific requirements and challenges of each field.

\subsection{Intuitive introduction to physics priors in MIA:}
Physics information is essential in MIA, improving accuracy and reliability. This information e.g. MRI acquisition physics for data augmentation or signal evolution physics for motion artifact correction, ensures better performance of PIMIA tasks. Additionally, PINNs model fluid dynamics and hemodynamics, enhancing predictive modelling. By incorporating physics into tasks like MRI reconstruction and CT artifact reduction, medical analyses become more robust and precise.
\begin{figure}[h]
    \centering
    \includegraphics[width=\linewidth]{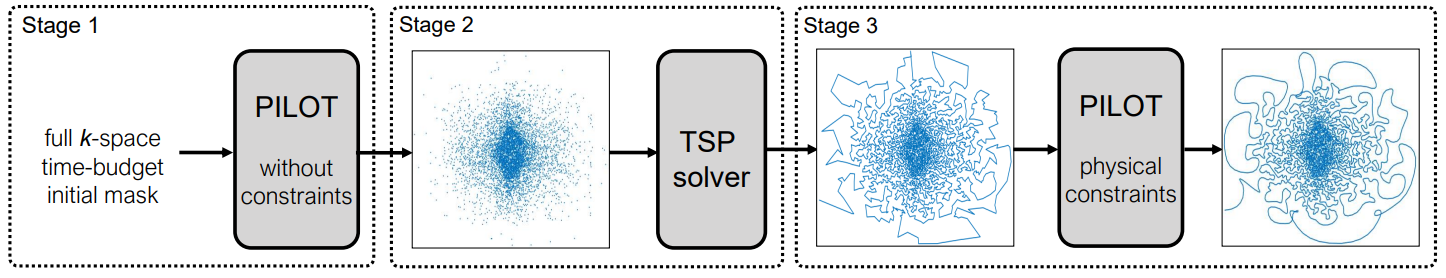}
    \caption{The three-stage process begins with the collection of random points in k-space, which represents spatial frequency information. These points are gathered without hardware constraints. Subsequently, a TSP (Travelling Salesman Problem) solver is employed to arrange these points into a path that minimizes distance. This optimized trajectory undergoes refinement during training to accommodate machine/ hardware constraints. This enhances MRI performance by improving data collection efficiency and image quality while ensuring operational feasibility.}
    \label{fig:intuit_example}
\end{figure}

We now discuss an intuitive example. \cite{weiss2019pilot} (see Fig:~\ref{fig:intuit_example}) optimize MRI trajectory by integrating physics constraints and solving the Traveling Salesman Problem (TSP), for accelerated MRI. This method ensures trajectories adhere to hardware limitations including gradient coil capacities and slew rates. By tackling TSP, PILOT-TSP reorders trajectory points to minimize path length while respecting physical constraints, and refining trajectory designs. This process optimizes MRI data acquisition, improving image quality and efficiency, particularly in single-shot scenarios. Through PILOT-TSP, the training pipeline effectively combines physics principles with computational techniques to advance MRI trajectory design, addressing challenges in accelerated imaging.

In other examples \cite{poirot2019physics} the authors employ DECT attenuation physics to facilitate material decomposition in brain imaging, using a custom ResNet-based architecture to harness the distinct attenuation properties at different energy levels in DECT scans. Similarly, \cite{eichhorn2024physics} uses signal evolution physics to correct motion artifacts in GRE MRI of the brain, with a custom MLP-based architecture designed to account for patient movement. Lastly, \cite{kissas2020machine} predict arterial flow in the cardiovascular system using 4D flow MRI data. These networks are trained to predict flow and pressure wave propagation, guided by conservation laws. Unlike traditional models, this method bypasses complex pre-processing and boundary conditions. Instead, it directly utilizes noisy clinical data, producing physically consistent predictions without the need for conventional simulators.

\subsection{Physics prior categories with examples (Taxonomy I):}
Depending on the source and form of the physics prior or information being incorporated in MIA tasks, we have categorised them into three major groups, which are elaborated with examples as follows.

\subsubsection{Medical Imaging Physics:} \label{sec:MedImg}
This category presents such works that have typically incorporated physics information and parameters derived from imaging systems and processes into medical analysis tasks. \cite{poirot2019physics} utilized DECT attenuation physics for material decomposition in brain imaging, while \cite{eichhorn2024physics}) corrected motion artifacts in GRE MRI brain scans using signal evolution physics. \cite{zhu2023physics} reduced metal artifacts in CT imaging with a beam hardening correction model.\cite{borges2024acquisition} and \cite{leung2020physics} applied MRI acquisition physics and PET modelling physics for data augmentation. \cite{jiang2021label} reconstructed heart image sequences with cardiac electrical activity and surface voltage data within a GCNN framework.
\subsubsection{Process dynamics and Equations:}\label{sec:DynEqn}
Here we present works that have primarily physics into medical analysis using Physics-Informed Neural Networks (PINNs) and similar methods, enhancing predictive modelling, imaging, and reconstruction. Works of \cite{halder2023mri}, \cite{zapf2022investigating}, and \cite{herrero2022ep} employ PINNs to incorporate fluid dynamics, 4D PDEs, and electrophysiological models, respectively. \cite{kissas2020machine} and \cite{sarabian2022physics} use conservation laws and hemodynamic equations, while \cite{van2022physics} and \cite{buoso2021personalising} leverage ODEs and cardiac mechanics constraints. Techniques in imaging and reconstruction, such as those in \cite{kamali2023elasticity} and \cite{zheng2024mr}, utilize elasticity theory and Helmholtz equations. Moreover, studies like \cite{zhang2023physics} and \cite{zhang2022physics} predict hemodynamics and musculoskeletal forces using computational fluid dynamics and physical laws. These integrations demonstrate the efficacy of physics-based constraints in improving accuracy and robustness in medical analyses.

\subsubsection{Others:} \label{sec:others}
Two key approaches have emerged for incorporating underlying physics into machine learning frameworks: Representation and Features, and Physical Information and Measurements. The former utilizes patterns, representations, or features to encode physics information, as exemplified by works integrating data confidence maps and periodic motion patterns into segmentation and registration tasks, respectively. while the latter involves directly leveraging physical principles and measurements, which are derived from biological study sources and domain experts.

\textit{Representation and Features:} These works use patterns, representations or features as a medium of incorporating underlying physics information into the learning process. (\cite{peiris2023uncertainty}) integrate data confidence maps into MRI/CT segmentation for pancreas and heart imaging. (\cite{chen2023cotrfuse}) utilize image data features for segmentation of skin and lung images in CXR and dermoscopy. (\cite{fechter2020one}) apply periodic motion patterns for deformable image registration in lung and heart CT/MRI. (\cite{liu2020rare}) use k-space artifact patterns to augment MRI denoising for liver imaging.

\textit{Physical information and measurements:} \cite{altaheri2022physics} utilize EEG signals in Brain-Computer Interface technology, inherently involving the physics of electrical brain activity and signal processing. \cite{frid2018gan} discuss the implicit physics underlying medical image generation, particularly in CT scans, governed by X-ray attenuation principles. \cite{shi2020knowledge} tackles data scarcity in medical imaging by leveraging domain knowledge from radiologists to synthesize high-quality images, especially for thyroid nodules.

\begin{figure*}[t]
    \centering
    \includegraphics[width=0.8\linewidth]{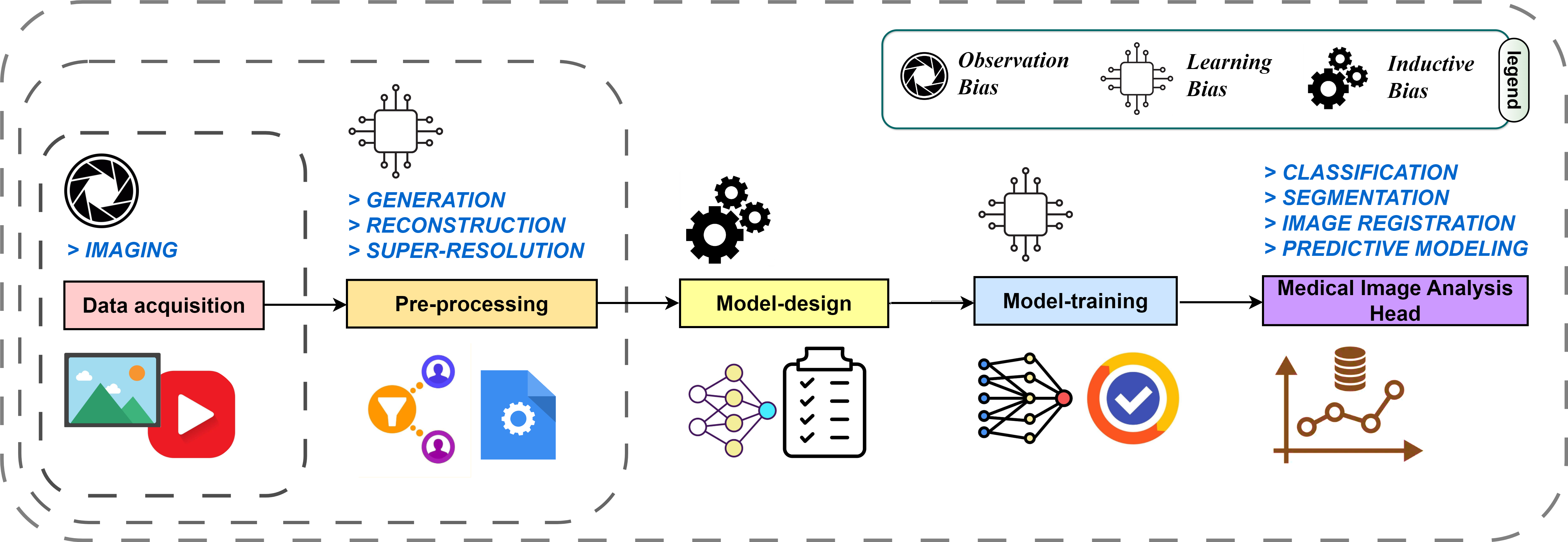} 
    \caption{This figure integrates the primary taxonomy (II) i.e. the MIA tasks, with the MIA pipeline backbone. It also introduces the bias-based scheme from PIML works and merges it with the pipeline. The illustration as a whole presents the typical locations of physics incorporation within the MIA pipeline concerning these biases.}
    \label{fig:cv_pipeline}
\end{figure*}


\subsection{Medical Image Analysis (MIA) tasks (Taxonomy II)}
Here we briefly introduce the different \textcolor{black}{MIA tasks} that we have used to organize the literature that we have reviewed.

\textit{Imaging} encompasses the process of capturing visual data using various sensors and techniques, which serves as the foundational input for PIMIA tasks. This process involves the acquisition, processing, and analysis of visual information to extract meaningful patterns and features for subsequent computational analysis.

\textit{Image generation} involves creating synthetic images using advanced algorithms, often leveraging deep learning frameworks such as Generative Adversarial Networks (GANs). This task is critical for producing realistic images from latent space representations, and facilitating applications in data augmentation, simulation, and creative design.

\textit{Inverse imaging} focuses on reconstructing original images from indirect or incomplete measurements by solving ill-posed problems. It employs mathematical and computational methods to infer high-quality images from degraded or noisy data, often utilizing regularization techniques to ensure stable and accurate reconstructions.

\textit{Image super-resolution} is the technique of enhancing the spatial resolution of an image by reconstructing high-frequency details from low-resolution inputs. This process involves upscaling the image using sophisticated algorithms that predict fine details and textures, thereby improving the visual quality and detail fidelity of the image.

\textit{Image reconstruction} refers to the process of forming a visual image from raw sensor data by applying computational techniques. This task involves the use of mathematical models to convert data from sensor measurements into interpretable images, typically through iterative algorithms that refine the image quality based on physical principles.

\textit{Image registration} is the process of aligning multiple images into a unified coordinate system. This involves geometrically transforming images to match a reference frame, ensuring spatial correspondence. It is essential for integrating data from different viewpoints, sensors, or time points, enabling coherent analysis and comparison.

\textit{Image segmentation and classification} involve partitioning an image into distinct regions (segmentation) and assigning labels to these regions (classification). These processes utilize machine learning and pattern recognition techniques to identify and categorize objects and structures within the image, facilitating detailed image analysis.

\textit{Predictive modeling} integrates physical laws and constraints into machine learning frameworks to enhance the accuracy and reliability of predictions. This approach combines data-driven methods with domain-specific physical knowledge, ensuring that model outputs adhere to known physical principles and exhibit realistic behaviour.

\subsection{Typical pipeline}
We introduce each stage of the MIA pipeline which serves as a typical backbone for MIA task and their connection with different types of biases:

\textit{Data acquisition:} In this stage, visual data such as images, videos, or sensor data is input to the MIA algorithm. Physics incorporation here falls under observational bias, where direct or simulated physics data is used. 

\textit{Pre-processing:}  Acquired visual data undergoes standardization to prepare it for MIA models. Physics-guided methods like super-resolution or image synthesis aid this stage, aligning with learning bias by enforcing physical constraints through regularization 

\textit{Model design:} This stage involves feature extraction and selecting/customizing model architectures. Physics is integrated through model design, enhancing feature extraction methods like using custom networks to extract transient features from images.

\textit{Model training:} MIA models are trained by optimizing parameters to minimize losses. Physics incorporation often occurs through loss functions that enforce physical laws like conservation equations. This approach fits under learning bias, ensuring models adhere to physical constraints during training.

\textit{MIA head:} This final component of the MIA pipeline encompasses modules that perform tasks such as prediction, segmentation, or reconstruction from visual data. These modules learn from the trained models to approximate functions or distributions by optimizing network parameters. The selection of loss functions directly impacts model efficiency, with common choices including cross-entropy and pixel-wise losses. Customized loss functions are also used to tailor training towards specific data characteristics and desired outcomes, enhancing model performance in targeted ways.

\begin{table*}[h!]
\caption{Categorization based on the type of physics prior used for physics incorporation in MIA tasks.}
\label{physics prior types}
\centering
\resizebox{\textwidth}{!}{
\begin{tabular}{|c|l|}
\hline
\multicolumn{2}{|c|}{\textbf{Categories of Physics information priors}}\\
\hline
A. & \textbf{Medical Imaging Physics }\\
\hline
& \cite{poirot2019physics}, \cite{eichhorn2024physics}, \cite{zhu2023physics}, \cite{halder2023mri}, 
\cite{kawahara2023mri},\cite{zhang2024phy} \\
& \cite{pan20232d}, \cite{borges2024acquisition}, \cite{leung2020physics}, \cite{tirindelli2021rethinking}, \cite{fok2023deep},\cite{borges2019physics}, \\
&  \cite{zimmermann2024pinqi}, \cite{shen2022geometry}, \cite{peng2022learning}, \cite{desai2021vortex}, \cite{weiss2019pilot}, \cite{jiang2021label} \\
\hline
B. & \textbf{Process dynamics and Equations} \\
\hline
& \cite{halder2023mri}, \cite{zapf2022investigating}, \cite{herrero2022ep}, \cite{kissas2020machine}, \cite{sarabian2022physics}, \cite{van2022physics}, \\
& \cite{buoso2021personalising}, \cite{lopez2023warppinn}, \cite{de2023spatio}, \cite{zhang2023physics}, \cite{zhang2022physics},
\cite{he2023optimization},\cite{kamali2023elasticity}\\
& \cite{kaandorp2021improved}, \cite{shone2023deep}, \cite{fathi2020super}, \cite{sautory2024unsupervised}, \cite{ragoza2023physics},\\ 
& \cite{zheng2024mr}, \cite{oh2020cycleqsm}, \cite{maul2024physics}, \cite{saba2022physics}, \cite{han2023diffeomorphic}, \cite{hunt2023fast}, \cite{min2023non},\cite{ruiz2022physics}\\
\hline
C. & \textbf{Others} \\
\hline
& \cite{peiris2023uncertainty}
\cite{chen2023cotrfuse}
\cite{fechter2020one}
\cite{liu2020rare}\\
& \cite{altaheri2022physics},
\cite{frid2018gan},
\cite{shi2020knowledge} \\
\hline
\end{tabular}
}
\end{table*}

\section{PIMIA Tasks: Detailed Survey} \label{sec:PIMIA_survey}



\subsection{Imaging}

Recent advancements in physics-guided machine learning models for imaging applications have demonstrated significant enhancements across various domains. 
\begin{figure}[h]
    \centering
    \includegraphics[width=\linewidth]{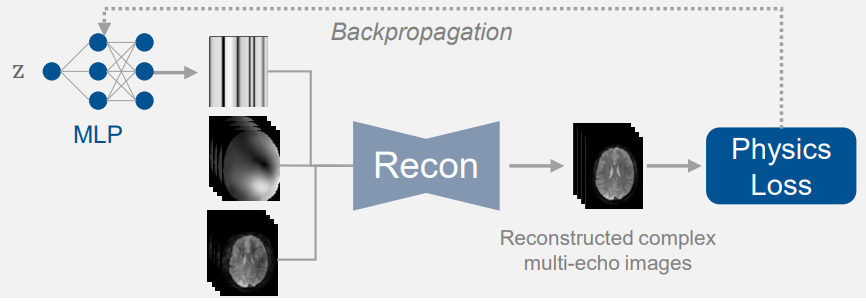}
    \caption{The figure shows a physics-informed process of PHIMO \cite{eichhorn2024physics}. Preceded by undersampled reconstruction, the above process trains a multi-layer perceptron (MLP) utilizing a physics loss, to predict exclusion masks for motion-corrupted lines. PHIMO eliminates intra-scan motion events, enabling precise correction for severe motion artifacts in multi-echo GRE MRI scans.}
    \label{fig:imaging_example}
\end{figure}
\cite{poirot2019physics} enhanced CT processing using a network based on the ResNet architecture that incorporates lookup virtual non-contrast (L-VNC) image data.

Other notable developments include \cite{eichhorn2024physics}, who introduced a motion correction method for quantitative MRI, leveraging physics-informed learning. It utilizes quantitative parameter estimation to exclude motion-corrupted k-space lines from reconstruction, particularly addressing challenges in T2* quantification from gradient echo MRI, enhancing image quality and reducing acquisition time for clinical applicability.
\cite{zhu2023physics}, presents a physics-informed sinogram completion (PISC) method for metal artifact reduction (MAR) in computed tomography (CT) imaging. Leveraging physical models, it corrects sinograms based on beam-hardening correction and normalized linear interpolation, effectively reducing metal artifacts while preserving structural details near metal implants.

The work by \cite{kamali2023elasticity}  utilizes PINNs to simultaneously reconstruct material parameters, including Young's modulus and Poisson's ratio, and stress distributions in linear elastic materials. It leverages the governing equations of linear elasticity, stress-strain relationships, and momentum balance equations. PINN integrates experimental measurements and finite element modelling to accurately estimate mechanical properties for clinical elasticity imaging applications.
%
In \cite{halder2023mri} the authors utilize real-time dynamic MRI and Physics-Informed Neural Networks (PINNs) to analyze esophageal bolus transport. It integrates detailed fluid dynamics and pressure-volume relationships, enhancing understanding of esophageal physiology. This approach offers improved diagnostics for swallowing disorders and advances computational modelling in gastroenterology.

\subsection{Image Generation}

Within the realm of MIAtasks focused on generation, 
In \cite{kawahara2023mri} the authors employs Generative Adversarial Networks (GANs) to generate images with consideration of MR properties. 
Additionally, \cite{pan20232d} focuses on medical imaging for data augmentation purposes, leveraging custom models and datasets like ACDC MRI and BTCV. Addressing Medical MRI analysis, \cite{borges2024acquisition} uses data augmentation techniques informed by UNet models and custom datasets, including ABIDE.

\begin{figure}[h]
    \centering
    \includegraphics[width=\linewidth]{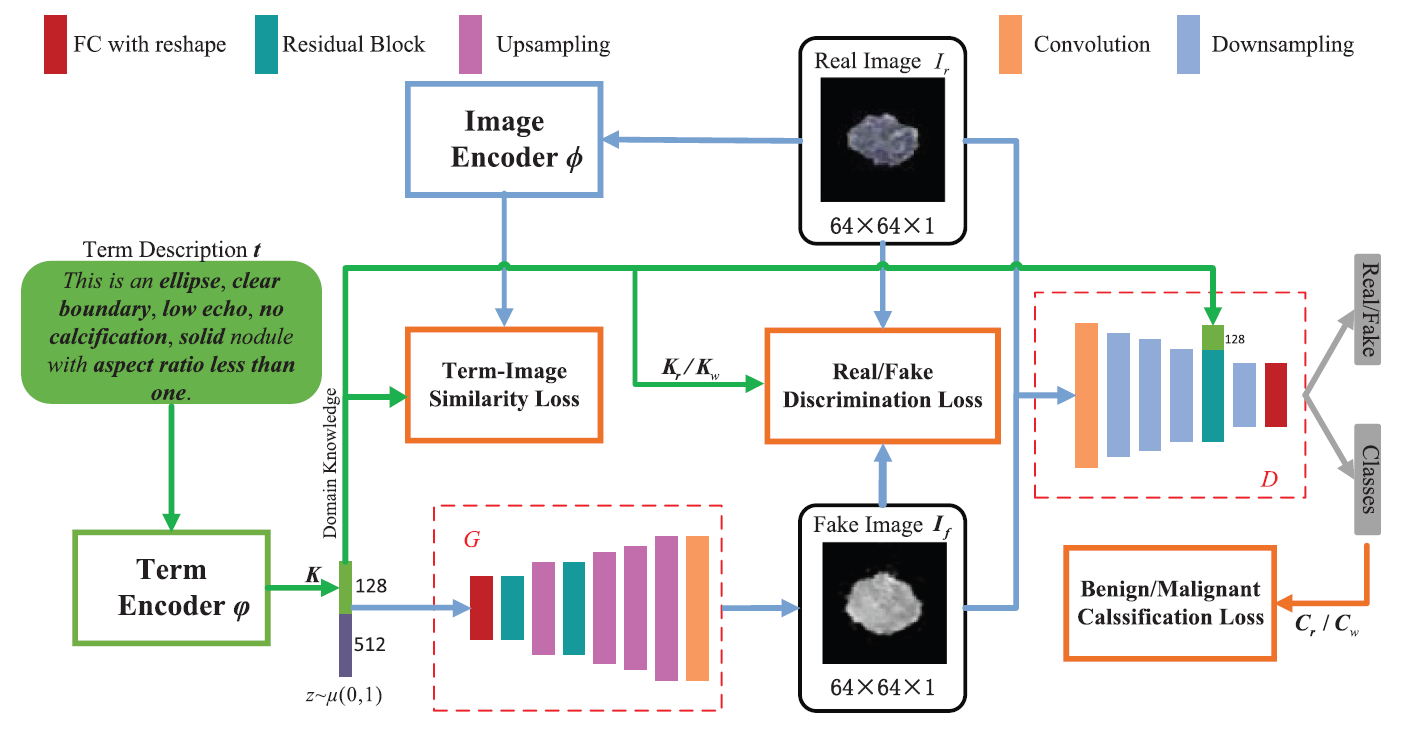}
   \caption{The figure shows the KACGAN architecture \cite{shi2020knowledge}. Here ultrasound images and their term descriptions are used as input. Term descriptions (t) are encoded into a 128-dimensional domain knowledge vector (K) through the term encoder $(\varphi)$ compression. This domain knowledge (K) is combined with a 512-dimensional noise vector (z) to generate fake images $(I_f)$ using the generator network (G). The discriminator network (D) integrates this domain knowledge for better classification. Additionally, an image encoder ensures the similarity between domain knowledge and the ultrasound images. This process incorporates physics into data augmentation through domain knowledge and word embedding, enhancing the learning and generation of high-quality images.}
    \label{fig:generation_example}
\end{figure}

%
\cite{leung2020physics} explores the integration of physics in PET imaging, focusing on the simulation of radioactive decay, photon interactions, detector responses, system blurring, and noise characterization. These simulations, crucial for realistic image generation, enhance the training and validation of neural networks (modified UNet) in medical imaging applications.
In the work by \cite{shi2020knowledge} the authors discuss a method employing an Auxiliary Classifier Generative Adversarial Network (ACGAN) to synthesize augmented medical images. By conditioning image synthesis on semantic labels, it enhances image diversity and realism for medical analysis. This approach leverages domain knowledge to generate synthetic data crucial for training machine learning models in medical imaging.

In their research \cite{frid2018gan} presents a method using Generative Adversarial Networks (GANs) to generate synthetic medical images, aiding in data augmentation for CNN-based liver lesion classification. GAN-synthesized images improve classification accuracy, addressing challenges of limited annotated datasets in medical imaging, and potentially enhancing diagnostic support for radiologists.
\cite{tirindelli2021rethinking} introduces a novel approach to ultrasound (US) data augmentation by integrating physics-inspired transformations. It proposes deformations, reverberations, and Signal-to-Noise Ratio adjustments, tailored to the principles of US imaging. These techniques aim to generate realistic variations in US images, enhancing their suitability for deep learning-based medical applications.
\cite{zhang2024phy} demonstrated a physics-guided diffusion model (Phy-Diff) for enhancing diffusion MRI (dMRI) synthesis. Phy-Diff integrates dMRI physics into noise evolution and employs query-based mapping, alongside the XTRACT atlas for anatomical details. It aims to improve dMRI quality through principled noise management and anatomically accurate synthesis.
In the work by \cite{momeni2021synthetic}, the authors generate synthetic microbleeds (sCMB) to enhance the training of neural networks for detecting cerebral microbleeds (CMB) in MRI scans. Using Gaussian modeling, it simulates diverse CMB characteristics, improving classifier performance without extensive ground truth. Physics-guided by MRI properties, sCMB mimics real lesions effectively for robust training.

\subsection{Inverse Imaging }
Physics-informed inverse imaging techniques leverage deep learning models incorporating physical principles to enhance medical imaging. For instance, 4D Flow MRI super-resolution integrates fluid dynamics through Physics-Informed Neural Networks (PINNs), while specialized architectures like SR UNet improve X-ray image fidelity. PINNs also aid in MRI reconstruction by embedding Navier-Stokes equations, optimizing k-space trajectories, and employing physical variables for accurate, high-resolution images. These methods, spanning various imaging modalities, ensure physically plausible outcomes, enhancing image quality and diagnostic capabilities in biomedical applications.

\subsubsection{Image Superresolution}

\begin{figure}[h]
    \centering
    \includegraphics[width=\linewidth]{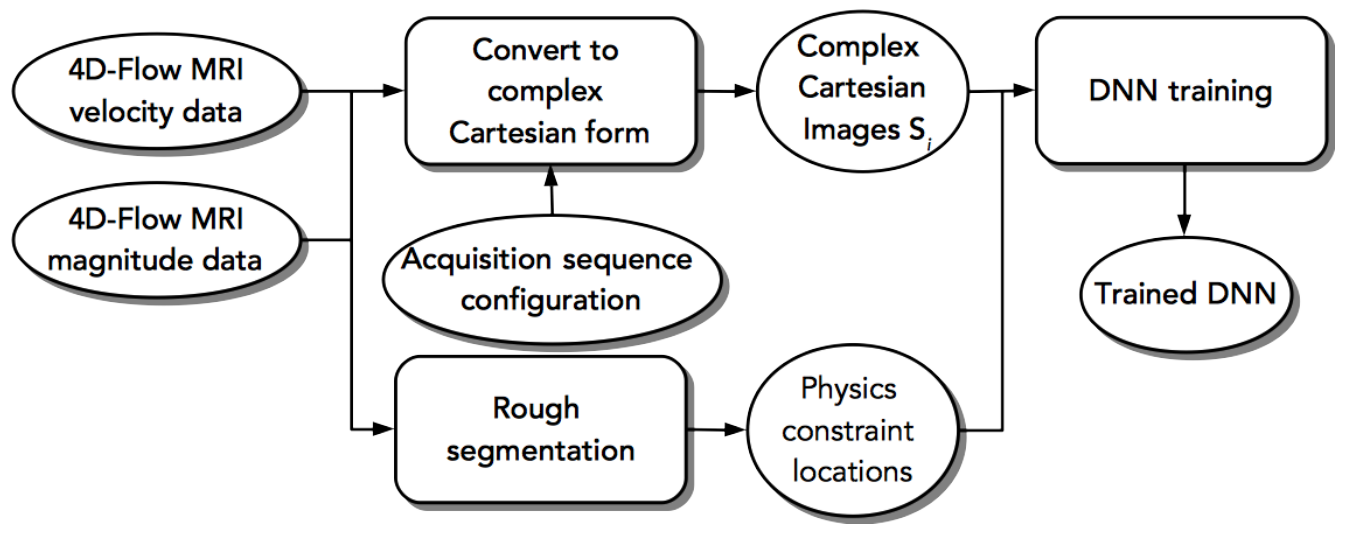}
   \caption{Data driven super-resolution and denoising of 4D-Flow MRI \cite{fathi2020super} . In the presented training pipeline flow velocities, pressure, and MRI image magnitude are modelled using a patient-specific deep neural network (DNN). The DNN is trained with 4D-Flow MRI images, incorporating fluid flow physics through regularization.}
   \label{fig:_example}
\end{figure}   

In the super-resolution task for medical image 
\cite{fathi2020super} focused on 4D Flow MRI, aiming to improve spatial resolution and noise reduction through super-resolution techniques, leveraging physics-informed deep neural networks to incorporate principles of fluid dynamics and mass conservation. Another paper discusses enhancing the quality of X-ray images through super-resolution, employing a specialized SR UNet architecture (referred to as SRUNK), which likely incorporates modulation transfer function kernels to achieve higher image fidelity \cite{fok2023deep}. 

Engaging in 4D Flow MRI, one study specifically targets super-resolution to enhance image quality, utilizing Physics-Informed Neural Networks (PINN) that integrate Navier-Stokes equations to inform the super-resolution process accurately \cite{shone2023deep}. 

Another study focuses on biomedical imaging, particularly MRI, aiming to reduce scan times through super-resolution techniques by employing SRGAN, which incorporates physical variables to ensure that the generated high-resolution images are physically plausible and of high quality \cite{chen2023physics}.

\subsubsection{ Image Reconstruction}

\begin{figure}[h]
    \centering
    \includegraphics[width=\linewidth]{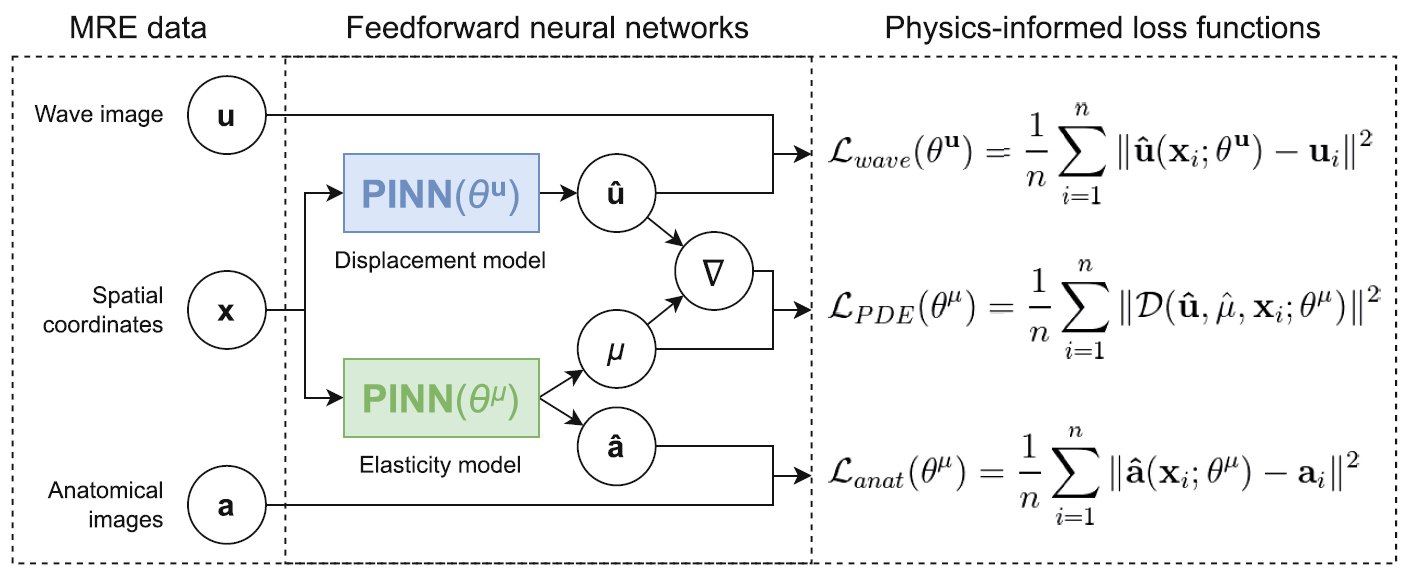}
   \caption{The pipeline from \cite{ragoza2023physics} employ physics-informed neural networks (PINNs) for the inverse problem, encoding displacement field $u(\mathbf{x})$ and elastic modulus $\mu(\mathbf{x})$ while adhering to a PDE. Using a dual-network strategy, one PINN fits wave images, minimizing ``wave loss" ($\mathcal{L}_{\text{wave}}$), while the other minimizes a PDE residual, termed ``PDE loss" ($\mathcal{L}_{\text{PDE}}$). An ``anatomical loss" function ($\mathcal{L}_{\text{anat}}$) refines predictions. Loss weights ($\lambda_{\text{wave}}$ and $\lambda_{\text{PDE}}$) fine-tune contributions during training for precise reconstruction.
 }
 \label{fig:reconstruction_example}
\end{figure}     

In medical image reconstruction tasks, 
in the work by \cite{burns2023untrained}, the authors focuses on super-resolution (SR) image reconstruction in the field of microscopy, a key area of interest in medical imaging, where enhancing the resolution of biological images is very important for better analysis and diagnosis.
%
\cite{zimmermann2024pinqi} introduced PINQI, a novel physics-informed neural network for quantitative MRI, integrates signal and acquisition models with learned regularization. It employs unrolled optimization for accurate parameter mapping, significantly improving performance on highly undersampled T1-mapping tasks using synthetic and real-world data.
Research by \cite{liu2020rare} introduced a RARE framework, enhancing MRI reconstruction by using priors from artifact-removal CNNs trained on undersampled data. It outperforms traditional denoising approaches, demonstrating effective 4D MRI reconstruction without fully-sampled groundtruth, validated on both simulated and experimental data.

\cite{peng2022learning} introduces a novel framework for accelerating Magnetic Resonance Imaging (MRI) by optimizing k-space sampling trajectories using a neural Ordinary Differential Equation (ODE). It frames k-space sampling as a dynamic system, incorporating MRI physics constraints, and uses a neural ODE to approximate trajectory dynamics, ensuring practical MRI acquisition.
The work study by \cite{oh2020cycleqsm} proposes an unsupervised deep learning method, leveraging physics-informed cycleGAN, to reconstruct quantitative susceptibility maps (QSM) from MRI phase images. By exploiting the known dipole kernel, it simplifies the architecture, providing faster and more stable training. This approach enhances QSM reconstruction accuracy, addressing the limitations of traditional supervised methods.

In thier work, \cite{jiang2021label} presents a novel approach to image reconstruction, integrating physics principles into a neural network framework. By leveraging electromagnetic theory, it models the relationship between cardiac electrical excitation and surface voltage signals. This physics-informed method improves reconstruction accuracy, offering insights into cardiac activity without requiring labelled data.
\cite{zheng2024mr} presents a novel approach, PIN-wEPT, for reconstructing electrical properties (EPs) in the human brain using magnetic resonance electrical properties tomography (MREPT). It integrates physics-informed neural networks with the Helmholtz equation, eliminating the need for specialized hardware configurations and providing accurate EPs distribution based on water content maps.

\cite{ragoza2023physics} demonstrates a novel approach for tissue elasticity reconstruction in Magnetic Resonance Elastography (MRE) using PINNs. Encoding physical principles through PINNs solves the inverse problem of tissue stiffness estimation without numerical differentiation, facilitating more robust and accurate elasticity mapping.
The research by \cite{shen2022geometry} introduces a novel approach, GIIR, for ultra-sparse 3D tomographic image reconstruction, integrating geometric priors of imaging systems. Utilizing a custom 3D-Net and geometric back-projection operator, it bridges 2D and 3D image domains, enhancing image reconstruction by leveraging physics-guided information, and promising advancements in biomedical imaging.

\cite{maul2024physics} introduces a method for reconstructing time-resolved contrast agent concentrations in angiography. Leveraging physics principles of blood flow and contrast agent transport, it trains a neural network using data from computational fluid dynamics simulations. This aids in efficiently approximating artifact-free reconstructions, improving angiographic image quality.
The work by \cite{desai2021vortex} proposed  VORTEX, a method for accelerated MRI reconstruction using physics-driven data augmentation. Leveraging knowledge of MRI physics, it augments MRI data with realistic noise and motion, enhancing robustness to distribution drifts. Composing image-based and physics-driven augmentations enables improved data efficiency and reconstruction quality in clinically relevant scenarios.

In \cite{weiss2019pilot}, the authors introduced PILOT (Physics-Informed Learned Optimized Trajectories), a novel approach to MRI acquisition and reconstruction. PILOT jointly optimizes k-space trajectories and image quality, integrating physical constraints on gradient coils and slew rates. It uses deep learning to design efficient acquisition schemes, enhancing MRI speed and quality.
\cite{saba2022physics} introduces a novel approach to optical diffraction tomography (ODT) by employing physics-informed neural networks (PINNs). Named MaxwellNet, it reconstructs three-dimensional refractive index distributions of biological samples, integrating physical principles directly into the reconstruction process. This technique enhances accuracy and efficiency while ensuring adherence to physics-based constraints.


\subsection{Image Registration}

\begin{figure}[h]
    \centering
    \includegraphics[width=\linewidth]{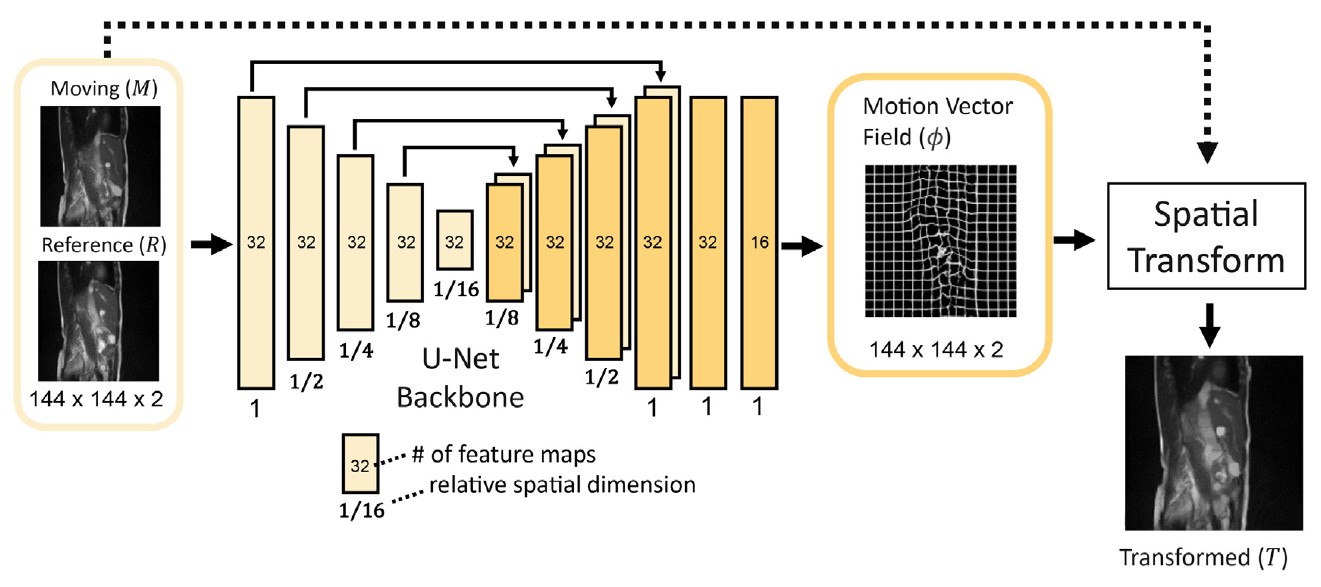}
    \caption{In \cite{hunt2023fast} the UNet-based model takes a reference and a moving image from cine MRI as inputs to generate a motion vector field (MVF), which aligns the images. This process accounts for complex patient motion during radiation therapy by capturing physical deformations in real time. By integrating these physics-informed transformations, the model enhances the accuracy and efficiency of image registration crucial for precise tumour targeting and dose delivery.}
    \label{fig:registration_example}
\end{figure}

\cite{fechter2020one} presents a novel approach for medical image registration, crucial for tasks like tracking organ motion. Utilizing deep learning and one-shot learning techniques, combined with physics-informed methodologies, it achieves accurate registration with minimal training data. The proposed algorithm demonstrates superior performance, especially for 4D datasets capturing periodic motion.

In their work \cite{min2023non} integrates medical image registration with linear elasticity equations, enabling precise alignment of images. By formulating the problem as a partial differential equation (PDE) system, it optimizes image transformation while preserving physical properties. This approach enhances accuracy and reliability in medical imaging applications, facilitating more effective diagnosis and treatment planning.
\cite{hunt2023fast} introduced a deep learning (DL) model for fast deformable image registration using 2D sagittal cine magnetic resonance imaging (MRI) acquired during radiation therapy. It leverages principles of MRI physics, respiratory motion, and real-time MRI-guided radiation therapy to develop a novel solution for real-time motion estimation in oncological treatments.

In their work \cite{he2023optimization} explores optimizing surface mesh generation parameters for biomechanical-model-based deformable image registration (BM-DIR) in liver and lung CT images. Incorporating physics principles, it refines boundary conditions for accurate image registration, vital for tasks like tumour tracking in radiation therapy. Physics-based modelling enhances precision in MIA and treatment planning.
\cite{han2023diffeomorphic} introduces DNVF, a diffeomorphic image registration method that incorporates physics-guided information using a neural velocity field represented by an MLP. This approach enables the precise alignment of medical images by modelling complex deformations, providing flexibility for optimization, and preserving desirable diffeomorphic properties.
%
%
%

\subsection{Image Segmentation \& Classification}

\begin{figure}[h]
    \centering
    \includegraphics[width=\linewidth]{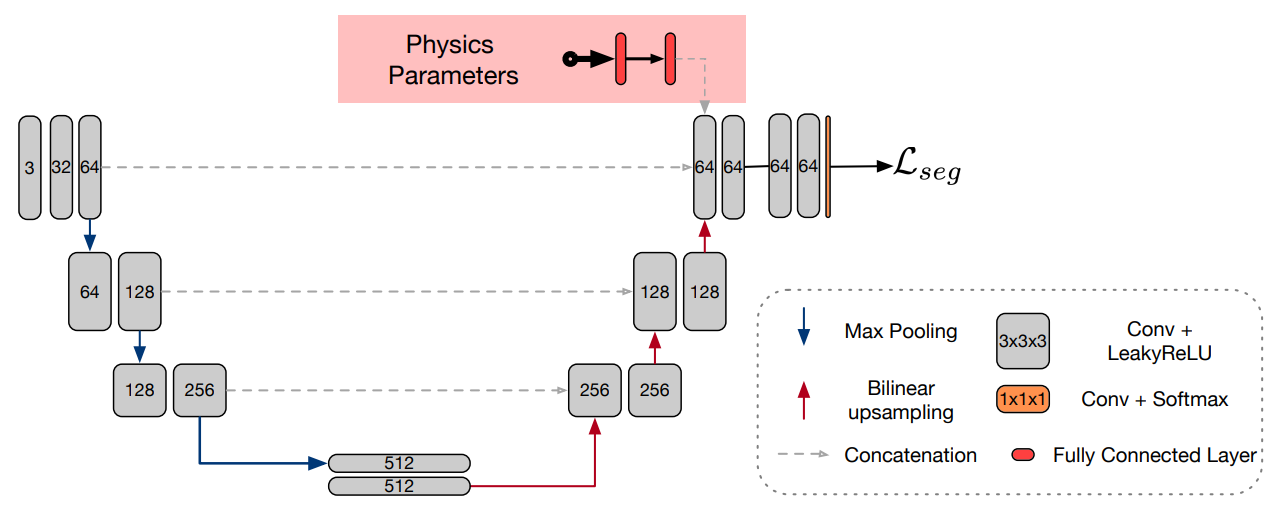}
   \caption{Physics informed segmentation network from \cite{borges2019physics}, the pink box constitute the novel contribution. Here physics parameters are integrated into a 3D U-Net for MRI segmentation. A dedicated "physics branch" with two fully connected layers incorporates N-dimensional vectors of physics parameters. These parameters, alongside their negative exponentiation, are concatenated into the network. The objective is to mitigate imaging parameter variability in MRI segmentations, aiming for consistency across acquisitions. }
   \label{fig:segmentation_example}
\end{figure}

Te work by \cite{borges2019physics} combines multiparametric MRI-based simulations with physics-informed CNNs, and demonstrates robust segmentation performance across varied image acquisition conditions. The methodology involves simulating MRI sequences and integrating sequence parameters into the CNN architecture to improve segmentation consistency.
\cite{chen2023cotrfuse} explores CoTrFuse, a novel network combining CNN and Transformer architectures for medical image segmentation. Leveraging attention mechanisms, it extracts both local and global information, improving segmentation accuracy.
In \cite{peiris2023uncertainty} the authors present Co-BioNet, a dual-view framework using adversarial learning for semi-supervised medical image segmentation. By leveraging labelled and unlabelled data, it outperforms existing methods, offering robustness and scalability, and improving multi-view learning.
\cite{altaheri2022physics} focuses on enhancing brain-computer interface capabilities for motor imagery classification, utilizing a custom model to analyze electroencephalogram (EEG) input data.

\subsection{Physics Informed Predictive Modeling}

In the domain of MIAtasks focused on prediction, \cite{kissas2020machine} engages in predicting arterial blood pressure within cardiovascular flow modelling, which holds significance for comprehending heart function and associated diseases. Additionally, \cite{zapf2022investigating} delves into estimating diffusion coefficients, an important endeavour contributing to our understanding of various medical and biological processes. 
Furthermore, \cite{sarabian2022physics} addresses brain hemodynamic prediction, enriching our comprehension of cerebral blood flow and its implications in both health and disease. In the context of electrophysiological parameter estimation, \cite{herrero2022ep} is instrumental, particularly in cardiac health and diagnostics.
The work by \cite{herrero2022ep} focuses on electrophysiological parameter estimation, a key area in cardiac health and diagnostics.
\cite{sarabian2022physics} addresses brain hemodynamic prediction, contributing to our understanding of cerebral blood flow and its implications in health and disease. \cite{zapf2022investigating} worked on estimating diffusion coefficients, which is crucial in understanding various medical and biological processes.\\
In their research \cite{kaandorp2021improved} introduces IVIM-NET, a deep learning model for diffusion-weighted magnetic resonance imaging (DWI). IVIM-NET integrates physics-based constraints to predict IVIM parameters reflecting tissue microstructure and perfusion. By enforcing physical principles through a tailored loss function, it accurately estimates diffusion and perfusion characteristics from DWI data, aiding in non-invasive tissue characterization.
\begin{figure}[t]
    \centering
    \includegraphics[width=\linewidth]{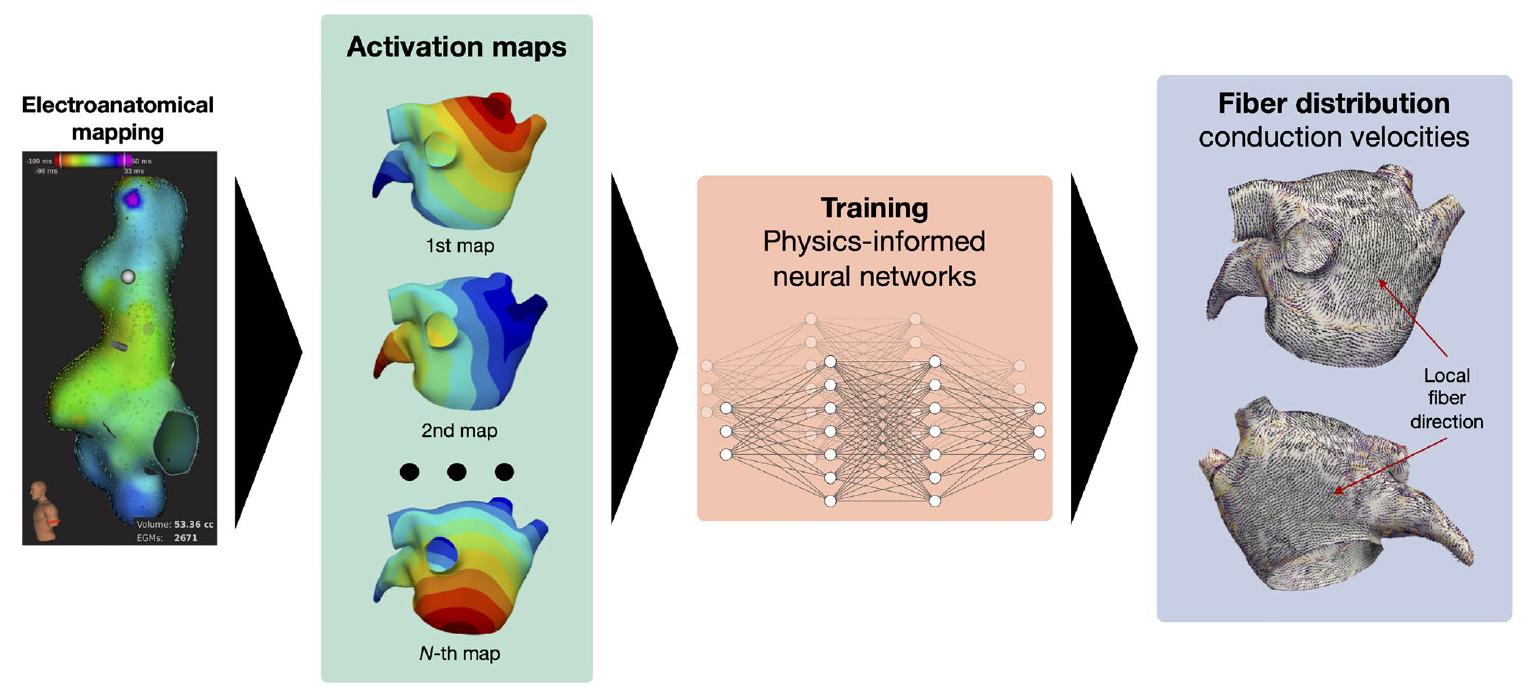}
   \caption{FiberNet \cite{ruiz2022physics} leverages physics-informed neural networks to translate multiple catheter recordings of cardiac electrical activation into a continuous estimation of cardiac fibre architecture and conduction velocity, crucial for predictive modelling based on the Eikonal equation. By solving an inverse problem, it deduces the conduction velocity tensor from sparse activation maps, enabling patient-specific models vital for personalized medicine.}
   \label{fig:prediction_example}
\end{figure} 

\cite{ruiz2022physics} proposes FiberNet, a Physics-Informed Neural Network (PINN) method to solve the inverse problem of identifying fibres in the heart using electroanatomical maps. By simultaneously fitting neural networks to maps and predicting conduction velocity tensors, it extends previous methods, demonstrating feasibility in clinical settings and validating through numerical experiments.
In their work \cite{zhang2022physics}  proposes a physics-informed deep learning framework for musculoskeletal modelling, integrating existing physics-based domain knowledge into data-driven models. It employs surface electromyography (sEMG) signals to predict muscle forces and joint kinematics, leveraging physics laws as soft constraints within a Convolutional Neural Network (CNN) architecture.

\cite{zhang2023physics} proposes a novel framework that integrates computational fluid dynamics (CFD) with physics-informed neural networks (PINNs) to predict hemodynamics in the cardiovascular system. By combining CFD simulations with deep learning, the framework enables real-time and noninvasive acquisition of hemodynamic parameters, facilitating diagnosis and treatment of cardiovascular diseases.
The work by \cite{de2023spatio} introduces SPPINN, a novel approach for CT perfusion analysis. SPPINN integrates physics principles into neural networks, enhancing accuracy in estimating cerebral perfusion parameters despite high noise levels. Leveraging spatio-temporal physics-informed learning, it offers a promising method for robust perfusion analysis in medical imaging applications.

\vspace{0.5cm}

\begin{table*}[h!]
\caption{
Organizing surveyed PIMIA papers into different MIA tasks, emphasizing unique aspects across categories: physics-guided operations, Region of Interest (ROI), imaging modality, training dataset, deep network architecture, and type of physics integration. Abbreviations used: Registr. - registration and Analy. - image analysis.(includes Segmentation and Classification), Super-res. - superesolution and Recon. - reconstruction.}
\label{table:main table}
\centering
\scalebox{0.48}{
\begin{tabular}{|c|l|l|l|l|l|l|l|}
\hline
& \textbf{Reference} & \textbf{Physics guided operation} & \textbf{ROI} & \textbf{Modality} & \textbf{Training dataset} & \textbf{Deep Network architecture} & \textbf{Type of Physics information (primary)} \\
\hline
\multirow{5}{*}{\rotatebox{270}{\textbf{Imaging}}} & 
\cite{poirot2019physics} & Material decomposition & Brain & DECT & Custom & Custom (ResNet based) & DECT attenuation physics \\
\cline{2-8}
& \cite{eichhorn2024physics} & Motion artifact correction & Brain & GRE MRI & Custom & Custom (MLP based) & Signal evolution physics \\
\cline{2-8}
& \cite{zhu2023physics} & Metal artifact reduction  & Diverse & CT & Custom & - & Beam hardening Correction Model \\
\cline{2-8}
& \cite{kamali2023elasticity} & Tumor diagnosis & Soft tissue & Elastography & Custom & PINN & Linear elastic theory, physical measurement \\
\cline{2-8}
& \cite{halder2023mri} & Esophageal disorder diagnosis & Esophagus & MRI & Custom & PINN & Fluid flow eqns., conservation laws \\
\hline
\multirow{8}{*}{\rotatebox{270}{\textbf{Generation}}} 
& \cite{kawahara2023mri} & Synthesis of FLAIR, DWI images & Brain & Synthetic & Custom & GAN & MR properties \\
\cline{2-8}
& \cite{pan20232d} & Image synthesis & Abdomino-Thoracic  & X-Ray, MRI, CT & ACDC MRI, BTCV etc. & Custom (MT-DDPM) & Diffusion process \\
\cline{2-8}
& \cite{borges2024acquisition} & Data augmentation & Brain & MRI & Custom, ABIDE & UNet & MRI acquisition physics \\
\cline{2-8}
& \cite{leung2020physics} & Data augmentation & Lung & PET & Custom & Custom (U-net based) & Physics of PET modelling \\
\cline{2-8}
& \cite{tirindelli2021rethinking} & Data augmentation & Spine & US & Custom & U-Net, DenseNet & US waves based physics information\\
\cline{2-8}
& \cite{shi2020knowledge} & Conditional image synthesis & Thyroid nodules & US & Custom & ACGAN & Domain knowledge (shape, Calcification etc.)\\
\cline{2-8}
& \cite{frid2018gan} & Liver lesion dataset synthesis & Liver & CT & Custom & Custom & Radiomic features of lesions \\
\cline{2-8}
& \cite{zhang2024phy} & raw dMRI image synthesis & Brain & diffusion MRI & HCP S1200 & HDiT & Diffusion physics\\
\cline{2-8}
& \cite{momeni2021synthetic} & Training data augmentation & Brain & SWI & AIBL\cite{ellis2009australian} & Custom  & Physical properties of Cerebral Microbleeds\\
\hline
\multirow{10}{*}{\rotatebox{270}{\textbf{Predictive Modeling}}} & \cite{zapf2022investigating} & Diffusion coefficient estimation & Brain & MRI & Custom & PINN & 4D PDE \\
\cline{2-8}
& \cite{herrero2022ep} & Electrophysiological & Heart & Optical mapping & Simulated cardiac EP data & PINN & PDE, ODE, IC and BC \\
\cline{2-8}
& \cite{kissas2020machine} & Predicting arterial & Artery & 4D flow MRI & Synthesized using DG solver & PINN & Conservation law constraints \\
\cline{2-8}
& \cite{sarabian2022physics} & Brain Hemodynamics & Brain & TCD US & Custom & PINN & 1D ROM PDE, Constraints \\
\cline{2-8}
& \cite{van2022physics} & MP MRI quantification & Heart & DCE-MRI & Custom & PINN & ODE residual loss \\
\cline{2-8}
& \cite{buoso2021personalising} & Cardiac mechanics simulation & Heart & MRI, CT & MMWHS & PINN & NN projection layer, cost function \\
\cline{2-8}
& \cite{lopez2023warppinn} & Cardiac strain estimation & Heart & MRI & Custom SSFP-MRI & PINN & Near-incompressibility of cardiac tissue \\
\cline{2-8}
& \cite{de2023spatio} & Estimate Cereberal perfusion parameters & Brain & CT perfusion &  ISLES 2018 & SPPINN & dynamics of CT perfusion \\
\cline{2-8}
& \cite{zhang2023physics} & Hemodynamics prediction & Cardiovascular & CTA & Custom & PointNet (PINN) & Comp. fluid dynamics, Navier-Stokes eqn.\\
\cline{2-8}
& \cite{zhang2022physics} & Muscle force and joint motion prediction  & Musculoskeletal & Surface-electromyography & Custom & CNN & Physics laws\\
\cline{2-8}
& \cite{ruiz2022physics} & Identification of fibers & Heart & Elctroanatomical (Elm.) map  & FiberNet (custom PINN) & Custom  & Physics info. extracted from Elm. maps\\
\cline{2-8}
& \cite{kaandorp2021improved} & Estimate diffusion-perfusion & Pancreas & DWI-MRI & NCT01995240, NCT01989000 & Custom IVIM-NET (PINN) & Physical principles of DWI-MRI\\
\hline
%
%
\multirow{6}{*}{\rotatebox{270}{\textbf{Inverse Imaging (Super-res.and Recon.)}}} & \cite{shone2023deep} & MRI Super-resolution & Heart & 4D-Flow MRI & Synthetic (CFD) & PINN & NS eqn, and symmetry constraints\\
\cline{2-8}
& \cite{fok2023deep} & CT super-resolution & Hand, wrist, elbow & CT (CBCT, MDCT) & Custom & Custom (based on UNet)  & Modulation Transfer Function kernels in CT\\
\cline{2-8}
& \cite{fathi2020super} & SR and denoising & Vascular system & 4D-Flow MRI & Custom & Custom & NS and conservation eqns., Fluid flow physics\\
\cline{2-8}
& \cite{sautory2024unsupervised} & Flow denoising, SR & Cardiovascular & MRV & synthetic (CFD) &  PINN & NS eqn., Newtonian fluid \\
\cline{2-8}
%
%
& \cite{zimmermann2024pinqi} & Quantitative MRI Reconstruction & Brain & MRI & Custom & Custom (PINQI) & MRI signal and acquisition model \\
\cline{2-8}
& \cite{liu2020rare} & Denoiser augmentation & Liver & MRI & Custom & Simplified DnCNN denoiser & k-space artifact pattern in 4D MR images \\
\cline{2-8}
& \cite{shen2022geometry} & 3D tomographic reconstruction & Lung & CT & LIDC, IDRI & 3D-Net (custom)  & imaging geometry priors\\
\cline{2-8}
& \cite{ragoza2023physics} & Tissue elasticity reconstruction & Liver & MRE(MRI) & Custom & PINN & PDE (Helmholtz equation)\\
\cline{2-8}
& \cite{zheng2024mr} & Reconstruction of electrical properties & Brain & MRI & Custom  & PINwEPT(PINN) & PDE (Helmholtz equation)\\
\cline{2-8}
& \cite{jiang2021label} & Image-sequence reconstruction & Heart & ECG & Custom & GCNN & Cardiac electrical activity and surface voltage \\
\cline{2-8}
& \cite{peng2022learning} & Accelerated MRI & Brain, Knee & MRI & fastMRI & PINN & MRI hardware constraints\\
\cline{2-8}
& \cite{oh2020cycleqsm} & QSM reconstruction & Brain & MRI & QSM 2016,19 challenges etc. & CycleQSM (custom PI-cycleGAN) & optimal transport theory \\
\cline{2-8}
& \cite{maul2024physics} & contrast agent conc. reconstr. & Cerebrovascular & DSA & Custom $+$ AneuX & CNN & CFD simulations\\
\cline{2-8}
& \cite{desai2021vortex} & accelerated MRI reconstr. & Knee & MRI & 3D FSE multi-coil Knee&  2D U-Net  & MRI signal and acquisition models\\
\cline{2-8}
& \cite{weiss2019pilot} & Accelerated MRI & Knee & MRI & NYU fastMRI & Custom (UNet based) & MRI hardware based constraints \\
\cline{2-8}
& \cite{saba2022physics} & reconstr. of 3D RID & Biological cells & ODT & Custom & PINN & PDE (Maxwell's equation) \\
\hline
\multirow{4}{*}{\rotatebox{270}{\textbf{Registr.}}} & \cite{han2023diffeomorphic} & Diffeomorphic image registration & Brain & 3D MRI & OASIS, Mindboggle & Custom (FCN based) & diffeomorphic deformation of images \\
\cline{2-8}
& \cite{he2023optimization} & Deformable image registration & Lung, Liver & CT (4DCT) & Custom & Morfeus & biomechanical models \\
\cline{2-8}
& \cite{hunt2023fast} & Deformable image registration & Abdomino-Thoracic & MRI & CineMRI(Custom) & VoxelMorph & MRI signal and acquisition models \\
\cline{2-8}
& \cite{min2023non} & Model prostrate motion & Prostate & MRI, TRUS & Custom (on prostrate cancer biopsy) & PINN & Linear elasticity equations \\
\cline{2-8}
& \cite{fechter2020one} & Deformable image registration & Lung, Heart & CT, MRI & DirLab, Popi, Sunnybrook & Custom (UNet based) & Periodic motion pattern \\
\hline
\multirow {3}{*}{\rotatebox{270}{\textbf{Analy.}}} 
& \cite{altaheri2022physics} & Motor imagery classification & Brain & EEG & BCI-2a dataset & Custom & EEG input data \\
\cline{2-8}
& \cite{borges2019physics} & Brain MRI segmentation & Brain & MRI & Custom, SABRE subsets & 3D U-Net & Physics parameter as training input \\
\cline{2-8}
& \cite{chen2023cotrfuse} & CXR/ Dermoscopic Segmentation & Skin, Lung & CXR, Dermascopy & ISIC-2017, COVID-QU-Ex & Custom (CoTrFuse) & Image data features \\
\cline{2-8}
& \cite{peiris2023uncertainty} & MRI/ CT Segmentation & Pancreas, Heart & MRI, CT & NIH Pancreas CT, LA MRI etc. & Custom (Co-BioNet) & Data Confidence maps \\
\hline
\end{tabular}}
\end{table*}

\section{Quantitative study and insights}\label{sec:quant_study}

\subsection{Calculation of performance improvement score}
\begin{figure}[h!]
    \centering
    \includegraphics[width=\linewidth]{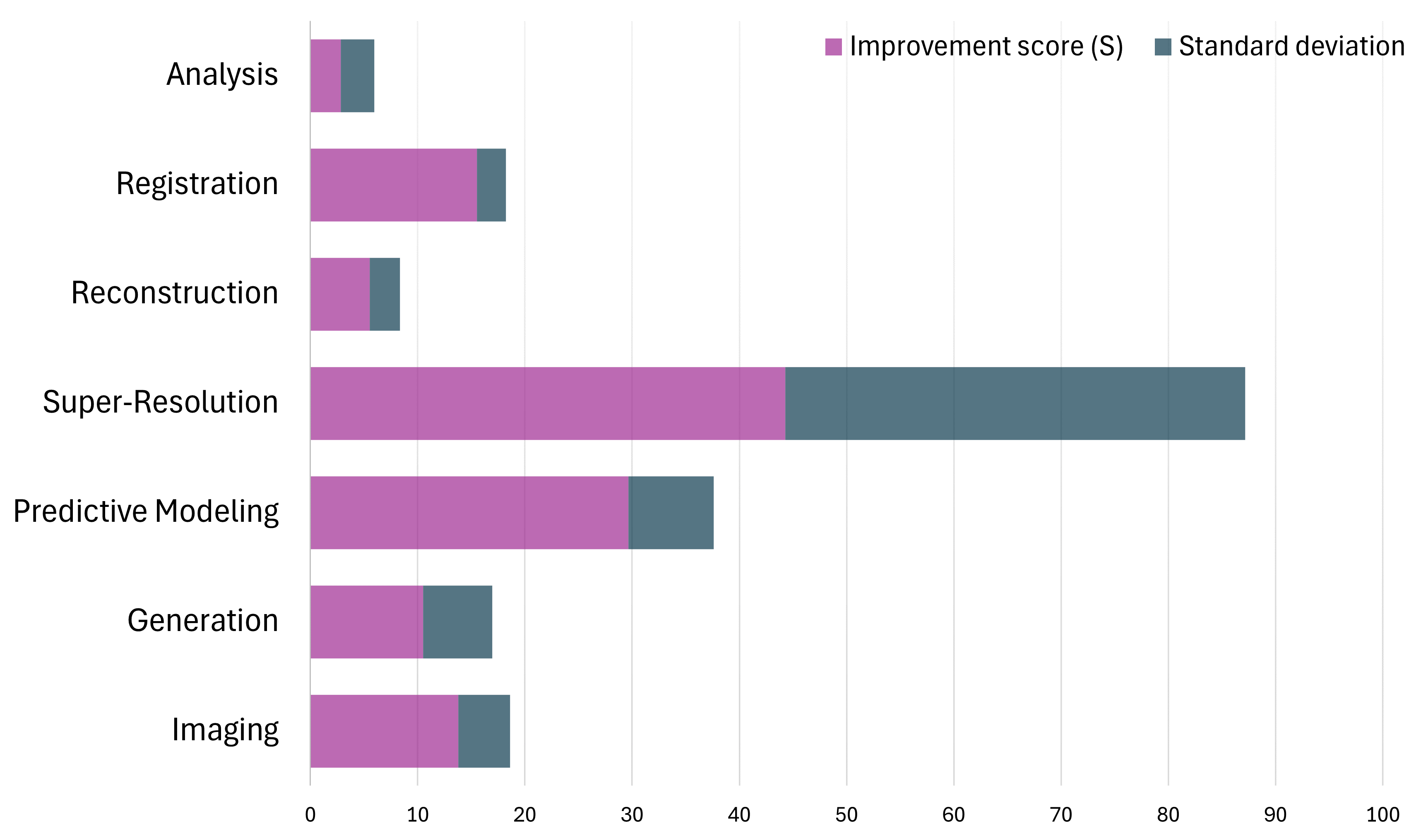}
    \caption{\textit{Representative comparison} of improvement in MIA task performance due to physics incorporation w.r.t. standard baselines. The performances are scored using custom performance measure i.e.\textit{Improvement score} ($\mathcal{S}_{Mean}$) and the corresponding standard deviation ($\mathcal{S}_{Std. dev.}$).The plot indicates the highest performance with significant variability in Super-resolution tasks, and the least performance improvement is observed in Analysis tasks, with registration and reconstruction tasks showing minimal variability. }
    \label{fig:perf_summary}
\end{figure}

We introduced a summary metric to show the improvement induced using physics information in PIMIA tasks. We selected a pair of relatively better-performing works from each task and represented them using our ``improvement score'' (see Fig.~\ref{fig:perf_summary}) for comparison. 

A custom score was necessary due to the use of different metrics, losses, baselines, ROIs, datasets, and imaging modalities, it was difficult to uniformly compare the performances even in a single task let alone across PIMIA tasks.
A plethora of metrics, from RMSE to F1-score, assess image quality, segmentation accuracy, registration precision, and predictive modelling efficacy. This diverse array spans from traditional measures like SSIM to indices like NMSE, SI and MSI, providing evaluation across various facets of PIMIA.

Following is the mathematical formulation of the improvement score ($\mathcal{S}$) calculation, considering, $\mathcal{P}$ as the performance achieved after incorporating physics information, $\mathcal{B}$ as the performance achieved without incorporating physics information. Additionally, $\mathcal{C}$ is the improvement score for a particular case, in a certain PIMIA task category.
Each $\mathcal{C}_{\text{\#case}}$ is calculated as the absolute percentage difference between $\mathcal{P}$ and $\mathcal{B}$, relative to $\mathcal{B}$. 

This is expressed as:

\[
\mathcal{C}_{\text{\#case}} = \left| \frac{(\mathcal{P} - \mathcal{B})}{\mathcal{B}} \times 100 \right|
\]

Then, the improvement score, denoted as $\mathcal{S}$, is calculated as the average of $\mathcal{C}_{1}$ and $\mathcal{C}_{2}$, where $\mathcal{C}_{1}$ and $\mathcal{C}_{2}$ represent the improvement scores for two different cases or published work in each PIMIA task category:

\[
\mathcal{S}_{Mean} = \frac{(\mathcal{C}_{1} + \mathcal{C}_{2})}{2}
\]

This formulation captures the relative improvement achieved by incorporating physics information across two cases, providing a standardized metric for comparison.
We also present the standard deviation of the above measure, which can be expressed as follows
\[
\mathcal{S}_{Std. Dev.}  = \sqrt{\frac{(\mathcal{C}_{1} - \mathcal{C}_{2})^2}{2}}
\]

The performance improvement score 
$\mathcal{S}_{Mean}$ serves as a key indicator of the enhancement achieved through physics-informed (PI) approaches. To simplify and encapsulate the essence of the PI-based improvement, we have applied a number of abstractions. When metrics or errors are reported as a mean and standard deviation in the source, only the mean value is used for comparison and score calculation. Furthermore, when performance data across multiple datasets or regions of interest (ROIs) is available, the baseline and improved scores with the greatest contrast are selected for analysis.

The performance data is based on the following works \cite{poirot2019physics,eichhorn2024physics} (imaging), \cite{leung2020physics,borges2024acquisition} (generation), \cite{van2022physics,kaandorp2021improved} (predictive-modeling), \cite{shone2023deep,fok2023deep} (Super-resolution), \cite{peng2022learning,desai2021vortex}(reconstruction), \cite{he2023optimization,min2023non} (registration) and \cite{borges2019physics,chen2023cotrfuse}(Segmentation/ Image analysis). Through these works we have considered the following errors and metrics, to measure the performance improvement: RMSE, MAE, DICE, NRMSE, SSIM and TRE.




\subsection{Performance improvement due to physics incorporation}

The comparative analysis of PIMIA tasks reveals significant variations in performance across different categories (see Fig:~\ref{fig:perf_summary}). The imaging task exhibits a notable 13.8\% average improvement. In generation tasks, the score improves by 10.495\%. Predictive modelling shows a substantial average improvement of 29.675\%. The inverse imaging task, specifically in super-resolution, demonstrates the most significant gains with an average improvement of 44.27\%. Conversely, tasks such as reconstruction and analysis display modest improvements, with average increases of 5.55\% and 2.85\%, respectively.

\subsection{Discussion: PIML Bias based taxonomy and PIMIA}

\textit{Observation bias} approaches typically use multi-modal data and DNNs to capture underlying physical principles from training data, sourced from direct observations, representations, or physics data. This bias is evident in works incorporating physical information, measurements, and feature-type physics priors (see Section~\ref{sec:others}). For example, \cite{peiris2023uncertainty} use data confidence maps for MRI/CT pancreas and heart segmentation, and \cite{chen2023cotrfuse} leverage image data features for segmenting skin and lung images in CXR and dermoscopy. \cite{fechter2020one} utilize periodic motion patterns for deformable image registration in lung and heart CT/MRI, while \cite{liu2020rare} enhance MRI denoising for liver imaging using k-space artifact patterns. Additionally, direct applications of physical principles are seen in \cite{altaheri2022physics} with EEG signals in Brain-Computer Interface technology reflecting electrical brain activity, \cite{frid2018gan} discussing the physics of X-ray attenuation in CT scan image generation, and \cite{shi2020knowledge} using radiologists' domain knowledge to synthesize high-quality images, addressing data scarcity in medical imaging.

A large number of the surveyed works have implemented physics priors through a \textit{Learning bias} type approach, i.e. enforcing prior knowledge/physics information through soft penalty constraints. Most of the approaches have directly used a PINN or a PINN-inspired methodology to implement the physics-based constraints. For example \cite{halder2023mri}, \cite{zapf2022investigating}, and \cite{herrero2022ep} employed PINNs to incorporate fluid dynamics, 4D PDEs, and electrophysiological models, respectively. Conservation laws and hemodynamic equations were utilized by \cite{kissas2020machine} and \cite{sarabian2022physics}, while \cite{van2022physics} and \cite{buoso2021personalising} leveraged ODEs and cardiac mechanics constraints. 

Concerning Inductive bias, where physics prior is incorporated through custom neural network-induced ``hard'' constraints. We could not find any individual work that has directly incorporated physics information through this approach.

\subsection{When to choose PIMIA over typical data-driven MIA?}

Incorporating physics information into data-driven models may become essential in certain specific scenarios. Typical data-driven approaches, while powerful in learning from diverse examples, require vast amounts of data, are time-consuming to train, and often lack theoretical guarantees, potentially disregarding physical principles. In the context of MIA physics-informed DNN models/ methods excel when data is limited, precision is important, and processing efficiency is essential. 

They are particularly effective in augmenting training and validation datasets by embedding known physical principles, such as MRI signal generation, which enables robust learning with limited data. This approach accelerates training, avoids local minima, and ensures solutions align with real-world physics, enhancing accuracy and reliability in medical imaging applications. These points are addressed in greater detail as follows:

\begin{itemize}
    \item [1)] \textit{Augmenting training and/ or validation dataset} 
    \newline
    Obtaining large annotated datasets can be challenging in medical imaging due to privacy concerns, high costs, and the need for expert annotations. For example, acquiring labelled MRI scans for rare diseases is difficult due to the low incidence of such cases. 
    Physics-informed models can compensate by embedding known physical principles, such as the physics of MRI signal generation, into the learning process. This allows them to perform reliably even with smaller datasets, maximizing the utility of limited data and ensuring robust and accurate models without needing extensive labelled images.

    \item [2)] \textit{Regularize learning of DNN models}\newline
    - \textit{Efficient training with smaller datasets}\newline
     Physics-informed (PI) models enhance training efficiency by incorporating physical principles, which act as constraints. This integration reduces the search space for model parameters, leading to faster convergence and improved performance even with smaller datasets. Consequently, these models require less data while maintaining high accuracy and reliability. For example, in MRI reconstruction, PI models use known MRI signal physics as constraints, which enables efficient learning from fewer scans. Thereby speeding up training and improving accuracy, requiring smaller datasets compared to purely data-driven models.
    
    - \textit{Helps avoid convergence to local minima.}\newline
      PI models combine the computational efficiency of physics-based algorithms with the adaptive learning capabilities of data-driven methods, thereby accelerating training and mitigating local minima issues. For example, in real-time MRI-guided surgery, physics-informed models offer rapid initial tissue structure approximations, reducing computational demands and ensuring timely, precision decisions crucial in clinical settings.
    
    - \textit{Making more physics-consistent solutions } \newline   
     Medical imaging techniques like MRI and CT scans are governed by complex physical principles which makes PI models particularly suited for these applications. For example, in cardiac MRI, understanding the blood flow dynamics and tissue properties is crucial for accurate diagnosis. Through incorporation of the physics of blood flow and tissue contrast, the models can better interpret the intricate details of cardiac images, leading to more accurate analyses. This allows for more effective identification of conditions such as heart valve defects or myocardial infarction, which purely data-driven methods might miss.
    
 \item [3)] \textit{Enables use of simpler models and architectures}\newline
- \textit{Smaller Models for Speed}\newline
Physics incorporation in models leads to simpler model architectures, thereby reducing the number of parameters and computational complexity. Resulting in comparatively smaller models that can process data faster, making them ideal for real-time applications and environments with limited computational resources, enhancing efficiency without sacrificing accuracy.

- \textit{Faster Convergence}\newline
PI models integrate physical laws and constraints, guiding the optimization process and reducing the search space for solutions. This leads to faster convergence during training, as the model quickly aligns with physically plausible outcomes. As a result, training times are shortened, and computational resources are conserved.

- \textit{Eradicate Impossible Solutions}\newline
By embedding physical principles directly into the model, physics-informed approaches eliminate impossible or non-physical solutions. This ensures that the model's predictions are consistent with real-world phenomena, enhancing reliability and trustworthiness in applications such as medical image analysis, where precision and realism are important.
\end{itemize}

\section{Current challenges and future research direction} \label{sec:Challenges}

\subsection{Challenges in Incorporating Physics into MIA}
The integration of machine learning and physics priors in MIA has ushered in significant advancements in the field of computational MRI, accurate tomographic reconstruction and better predictive models to name a few. Despite the progress, several open questions and specific challenges remain, which need to be addressed to further refine and enhance these approaches. The following discussion outlines the key issues and challenges faced in the application of physics-informed methods in MIA:

\begin{itemize}
\item [1)] \textit{Assessment of Overregularization Performance:}
The assessment of overregularization performance involves evaluating deep learning models that may suffer from overfitting due to excessive constraints during training. In clinical settings, these models can generate visually realistic but potentially misleading images, complicating artifact identification. This challenge is exacerbated by the presence of artifacts, where constraints imposed by prior information (physics priors) may inadvertently generate or obscure clinically significant features. Addressing this issue requires robust methods to quantify how these constraints influence model performance, ensuring maintained diagnostic accuracy amidst variable data quality in clinical practice.

\item [2)] \textit{Model Generalization:}
It's essential for ML models in MIA to adapt across diverse imaging equipment, sites, and populations. For example, variations in image acquisition, in terms of image quality, temporal resolution, Field of View (FOV) and patient positioning/ movement can significantly affect model efficacy. Integrating physics-based priors, which encompass anatomical knowledge,  imaging technique and acquisition metadata, is pivotal in addressing these variations. The key consideration lies in whether to develop versatile models using these priors for broad applicability or specialized models optimized for specific imaging scenarios to enhance diagnostic precision.

\item [3)] \textit{Prospective and Retrospective acceleration with physics priors:}
Medical imaging often utilizes retrospectively accelerated acquisitions for faster image generation, but this may overlook certain signal acquisition effects. Prospective acceleration, collecting data in real-time, offers more comprehensive information but presents challenges in transitioning from retrospective methods. Incorporating physics priors ensures that computational models effectively account for the nuances and complexities of the imaging process. The challenge lies in efficiently incorporating the complex physics of imaging systems e.g. MRI scanners in computational imaging methods. 

\item [4)] \textit{Use of processed data:}

Using processed images instead of raw data in MIA can sometimes lead to overly optimistic findings. This happens because processed images (e.g. DICOM) are altered representations of the original data (e.g. raw k-space data), potentially distorting the information in ways that may not accurately reflect real-world conditions. The challenge is in designing experiments that avoid such biases, ensuring that the findings derived from processed data align well with what would be observed in real-world scenarios, such as during patient scans. Adding physics information further complicates this alignment, requiring careful planning and refinement of experimental methods to maintain accuracy and reliability in MIA.
\end{itemize}

\subsection{Future directions}
Future directions include strengthening deep learning models across diverse imaging settings, refining uncertainty quantification in reconstruction, and integrating transformers with physics-based methods to optimize feature recognition and address data scarcity challenges in tomography.
\begin{enumerate}
    \item \textit{Enhancing Model Robustness Across Imaging Settings:}
          Strengthening deep learning models to withstand variability in imaging devices and settings through physics-based priors for reliable medical image analysis. E.g. enabling models to handle variations in MRI scanners by integrating prior knowledge of magnetic field strengths and imaging protocols.
          
    \item \textit{Improving Uncertainty Quantification in Reconstruction:}
          Refining algorithms to accurately quantify uncertainty in image reconstruction under diverse sampling patterns, leveraging physics-based insights. E.g. enhancing algorithms to accurately assess uncertainty in CT scans with different slice thicknesses, using physics-based insights on X-ray attenuation and reconstruction techniques.
          
    \item \textit{Advancing DNN Integration with Physics-Based Methods:}
        Combining \textit{transformers} (an advanced ML model architecture) with physics-based approaches can direct attention to critical features, boosting performance. Exploring self-supervised learning with physics-based methods can address challenges in data-driven tomography, particularly with limited paired training data.

\end{enumerate}



\section{Conclusions}\label{conclusion}
This paper introduces a state-of-the-art PIMIA paradigm that integrates data-driven methods with insights from physics and scientific principles. We present a unified taxonomy to classify PIMIA approaches based on the physics information, their representation and incorporation in MIA models. Our review covers a wide range of tasks, including imaging, generation, prediction, inverse imaging (super-resolution and reconstruction), registration, and image analysis (segmentation and classification). A comprehensive summary of the discussed papers is provided in a tabular format in Table~\ref{table:main table} to facilitate an understanding of how physics principles are integrated into MIA tasks.

The goal is to demystify the application of PIMIA methods across various MIA tasks, address current challenges, and encourage further research in this field.

\section*{Acknowledgments}
This research was partly supported by the Advance Queensland Industry Research Fellowship AQIRF024-2021RD4.

\bibliographystyle{model2-names.bst}\biboptions{authoryear}
\bibliography{PICV}

\end{document}